\documentclass[a4paper,11pt]{article}
\pdfoutput=1 

\usepackage{jheppub} 

\usepackage[T1]{fontenc} 

\usepackage{bm}

\newcommand{\la}{\langle}
\newcommand{\ra}{\rangle}
\newcommand{\nn}{\nonumber}
\renewcommand{\b}{\bm}
\renewcommand{\r}{\mathbf}
\renewcommand{\c}{\mathcal}
\renewcommand{\u}{\underline}

\newcommand{\fir}{h}
\newcommand{\fuv}{f}

\title{\boldmath Spontaneous Symmetry Breaking from an On-Shell Perspective}
\author[a]{Brad Bachu}
\affiliation[a]{Department of Physics, Princeton University\\NJ 08450, U.S.A}

\emailAdd{bbachu@princeton.edu}

\abstract{We show how the well known patterns of masses and interactions that arise from spontaneous symmetry breaking can be determined from an entirely on-shell perspective, that is, without reference to Lagrangians, gauge symmetries, or fields acquiring a vacuum expectation value. To do this, we review how consistent factorization of $2\rightarrow 2$ tree level scattering can lead to the familiar structures of Yang-Mills theories, and extend this to find structures of Yukawa theories. Considering only spins-$0$, $1/2$ and $1$ particles, we construct all the allowed on-shell UV amplitudes under a symmetry group $G$, and consider all the possible IR amplitudes. By demanding that on-shell IR amplitudes match onto on-shell UV amplitudes in the high energy limit, we reproduce the Higgs mechanism and generate masses for spins-$1/2$ and $1$, find that there is a subgroup $H \subseteq G$ in the IR, and other interesting relations. To highlight the results, we show the breaking pattern of the Standard Model $U(1)_{EM} \subset SU(2)_L \times U(1)_Y  $, along with the generation of the masses and interactions of the particles.}

\begin{document} 
\maketitle
\flushbottom

\section{Introduction}

On-shell techniques have proven themselves as not only an alternative method to compute scattering amplitudes more efficiently, but also as a tool to unlock underling structures of quantum theories. Most of the progress of on-shell techniques has been limited to amplitudes containing only massless particles. Though there has been progress with on-shell massive amplitudes \cite{Craig:2011ws,Kiermaier:2011cr,Conde:2016vxs,Conde:2016izb}, the techniques changed in a fundamental way when \cite{Arkani-Hamed:2017jhn} introduced spin-spinors for massive particles, natural counterparts to helicity-spinors for massless particles.
This has since opened the doors for a variety of massive on-shell works, including the computation of massive amplitudes \cite{Durieux:2020gip,Christensen:2018zcq,Ochirov:2018uyq,Wu:2021nmq,Chiodaroli:2022ssi}, effective field theories \cite{Liu:2023jbq,Balkin:2021dko,Durieux:2019eor,Aoude:2019tzn}, gauge invariance \cite{Liu:2022alx} and supersymmetry \cite{Herderschee:2019ofc}. 
In this paper, we continue these efforts and generalize the on-shell Higgs mechanism and spontaneous symmetry breaking, while finding interesting results along the way.

The Higgs mechanism and spontaneous symmetry breaking are traditionally described from a field theoretic perspective. Again, it was not until \cite{Arkani-Hamed:2017jhn} that it was highlighted how to achieve it on-shell. Further, the works of \cite{Bachu:2019ehv,Durieux:2019eor} showed how this could be applied to the electroweak symmetry breaking of the Standard Model (SM). In this work, we generalize some aspects of the Higgs mechanism and spontaneous symmetry breaking. In particular, we show how masses for spins $1/2$ and $1$ transforming under a general group $G$, can be generated from an entirely on-shell perspective, that is, without reference to Lagrangians, gauge symmetries, or fields acquiring a vacuum expectation value.

It is by now well known that the structure of Yang-Mill theories can be derived on-shell. To give mass to fermions without reference to Lagrangians, we use consistent factorization of $2\rightarrow 2$ tree level amplitudes to derive the structure of Yukawa theories under a general group $G$. Our strategy for generating masses for spins-$1/2$ and $1$ will be to demand consistency between a high energy theory (UV) and a low energy theory (IR). The consistency is implemented by imposing that the high energy limit of the IR matches onto the UV. As such we construct all the allowed on-shell UV amplitudes under a symmetry group $G$, and consider all the possible IR amplitudes. 
After we succeed in generating masses for a general theory, we will show how these results can be used to describe the process of electroweak symmetry breaking in the Standard Model. The nature of these results provide an opportunity for a pedagogical introduction.

This paper is structured as follows. We begin with a brief review of the little group and the construction of scattering amplitudes in Section \ref{sec:scattering_amps}, while deferring explicit details of helicity and spin spinors to Appendix~\ref{sec:kinematics}. We then use these ideas to construct three and four particle amplitudes in Section \ref{sec:MasslessThreeandFourParticleAmplitudes}, which are necessary for the structures of Yang-Mills and Yukawa theories. These two sections set the foundations for the UV-IR matching that we will perform. In Section \ref{sec:Non-AbelianHiggs}, which experts start at, we show how demanding an IR theory be consistent with a UV theory in the high energy limit allows us to discover the on-shell version of the Higgs Mechanism for spins-$1/2$ and $1$ particles. Lastly, in Section \ref{sec:StandardModel} we show how to interpret these results within the Standard Model.

\section{Scattering Amplitudes and the Little Group\label{sec:scattering_amps}}

In this section, we review the on-shell construction of amplitudes with helicity and spin spinors, and defer to \cite{Weinberg:1995mt,Arkani-Hamed:2017jhn,Liu:2022alx} for more details and related discussions. We start the discussion with the little group for massless and massive particles, along with their representations. We then review the properties of scattering amplitudes, and note that they must be Lorentz invariant and little group covariant, which motivates the introduction of helicity and spin spinors.

\subsection{The Little Group}

Following Wigner \cite{Wigner:1939cj}, we can think of particles as irreducible representations of the Poincare' group. We can diagonalize the translation operator with their momentum $p^\mu$, and label any other quantum numbers with $\sigma$. Using the reference momentum trick, we can write any momentum $p$ as a Lorentz transformation $L(p;k)$ acting on a reference momentum $k$ i.e., $p = L(p;k)k$. Assuming that we have unitary representations $U$ of elements of the Lorentz group $\Lambda$, we define one-particle states as
\begin{align}
    |p,\sigma\ra &= U(L(p;k))|k,\sigma\ra\,.
\end{align}
Now, $L(p;k)$ is not unique, as there are many Lorentz transformations that also leave $k$ invariant. The subgroup of the Lorentz group that leaves momentum invariant is referred to as the little group i.e., for $W$ an element of the little group, $Wk = k$. Then, the non-uniqueness of $L(p;k)$ is seen as the freedom to add little group transformations via $p = L(p;k)k = L^\prime(p;k)k$, for $L^\prime(p;k) = L(p;k)W$.

The little group plays an important role in characterizing how single particle states transform, so let us make the above statement more precise. First, consider a general Lorentz transformation $\Lambda$ acting on $p$. Using the reference momentum trick, we can write this as $\Lambda p = \Lambda L(p;k)k = L(\Lambda p;k)k$. Next, consider the identity $1 = L(\Lambda p;k) L^{-1}(\Lambda p;k)$ acting on $\Lambda p$,
\begin{align}
    1\,\Lambda p = L(\Lambda p;k) L^{-1}(\Lambda p;k)\Lambda L(p;k)k &= L(\Lambda p;k) L^{-1}(\Lambda p;k)L(\Lambda p;k)k \nonumber \\
   L(\Lambda p;k) W(\Lambda,p;k) k &= L(\Lambda p;k)k\,,
\end{align}
where we have identified $W(\Lambda,p;k) =L^{-1}(\Lambda p;k)\Lambda L(p;k)$ as an element of the little group! To act on the state $|k,\sigma\ra$, we simply need unitary representations, 
\begin{align}
    U(W(\Lambda,p;k)) |k,\sigma\ra &= D_{\sigma\sigma^\prime}(W(\Lambda,p;k)) |k,\sigma^\prime\ra \,,
\end{align}
where $D_{\sigma\sigma^\prime}(W)$ is a unitary representation of the little group.

Finally, consider the action of a general Lorentz transformation on the state $|p,\sigma\ra$. From the discussion above, we see that,
\begin{align}\label{eq:general:lt:particle}
    U(\Lambda)|p,\sigma\ra &= D_{\sigma\sigma^\prime}(W(\Lambda,p;k))|\Lambda p,\sigma^\prime\ra\,,
\end{align}
and so, we conclude that a single particle state is labelled by its momentum and transforms under some representation of the little group.

We end this subsection by elucidating the form of $D_{\sigma\sigma^\prime}(W)$. There is a clear distinction of the little group for massless and massive particles. For massless particles in four dimensions, the little group is $SO(2) = U(1)$, and representations of $U(1)$ are labeled by the integers $ n =  2h$, where $h$ is referred to as the particle's helicity. For massive particles in four dimensions, the little group is $SO(3) = SU(2)$, and finite dimensional irreducible representations of $SU(2)$ are labeled by  non-negative integers $n =2S+1$, where $S$ is referred to as spin. The simplicity introduced in \cite{Arkani-Hamed:2017jhn}, was to use symmetric tensor representations of $SU(2)$. That is, an irreducible spin-$S$ representation of $SU(2)$ corresponds to a fully symmetric rank $2S$ tensor.

With this in mind, we can revisit eq.~\eqref{eq:general:lt:particle} for massless and massive particles. Under a general Lorentz transformations, a massless particle with helicity-$h$ transforms as,
\begin{align}
    U(\Lambda)|p,h,\sigma\ra &= D_{\sigma\sigma^\prime}(W(\Lambda,p;k))|\Lambda p,h,\sigma^\prime\ra\,, 
\end{align}
where,
\begin{align}\label{eq:rep:littlegroup:massless}
    D_{\sigma\sigma^\prime}(W(\Lambda,p;k)) &= w_h \delta_{\sigma\sigma^\prime}\,,
\end{align}
for $w_h\in U(1)$. 
Under general Lorentz transformations, a massive particle with spin-$s$ transforms as
\begin{align}
    U(\Lambda)|p,\{I_1,\dots,I_{2s}\},\sigma\ra = D_{\sigma\sigma^\prime}(W(\Lambda,p;k)) |\Lambda p,\{J_1,\dots,J_{2s}\},\sigma^\prime\ra\,,
\end{align}
where
\begin{align}\label{eq:rep:littlegroup:massive}
    D_{\sigma\sigma^\prime}(W(\Lambda,p;k)) &= \delta_{\sigma\sigma^\prime} W_{I_1}^{\;\;J_1}\dots W_{I_{2s}}^{\;\;J_{2s}}\,,
\end{align}
for $W\in SU(2)$.

\subsection{Scattering Amplitudes}

Next, lets consider what this means when scattering $n$ particles, labelled by $|p_a,\rho_a\ra$, where $\rho$ represents additional quantum numbers needed to specify the particle.
As mentioned above, for massless particles of helicity-$h$, $\rho = (h,\sigma)$, and for massive particles of spin-$s$, $\rho = (\{I_1,\dots,I_{2s}\},\sigma)$.
Since we consider all particles outgoing, the scattering amplitude is defined as
\begin{align}
    \c{M}(p_1,\rho_1; \dots ; p_n,\rho_n) = _{\text{out}}\la p_1,\rho_1; \dots ; p_n,\rho_n | 0 \ra_{\text{in}} \,.
\end{align}
Poincare' invariance of the S-matrix implies,
\begin{align}
  \c{M}(p_1, \sigma_1; \dots; p_n, \sigma_n) &= \delta^4\left(\sum p_a\right)\c{A}(p_1, \sigma_1; \dots; p_n, \sigma_n)\,,
\end{align}
where we have used translation invariance to pull out a momentum conserving delta function.

To see the action of the little group in a scattering amplitude, let us consider the transformation law of $M(p_1, \sigma_1; \dots; p_n, \sigma_n)$ under a general Lorentz transformation $\Lambda$. Assuming that the asymptotic multi-particle states transform as a tensor product of one-particle states i.e. $|p_1,\rho_1; \dots ; p_n, \rho_n\ra = |p_1,\rho_1\ra \otimes \dots \otimes |p_n,\rho_n\ra$, so that $U(\Lambda)|p_1,\rho_1; \dots ; p_n, \rho_n\ra =U(\Lambda)|p_1,\rho_1\ra \otimes \dots \otimes U(\Lambda) |p_n,\rho_n\ra$, then
\begin{align}
    U(\Lambda) \c{A}(p_1, \sigma_1; \dots; p_n, \sigma_n) &=\left( \prod_{a}D_{\sigma_a\sigma_a^\prime}(W) \right)\c{A}(\Lambda p_1, \sigma_1; \dots; \Lambda p_n, \sigma_n) \,,
\end{align}
where $D_{\sigma_a\sigma_a^\prime}(W)$ takes the form of eq.~\eqref{eq:rep:littlegroup:massless} or eq.~\eqref{eq:rep:littlegroup:massive} if $a$ is a massless or massive particle respectively.

So, the $\c{A}(p_1,\sigma_1, \dots, p_n, \sigma_n)$ must be Lorentz invariant and covariant under the little group. With this in mind, we can further determine what form $\c{A}(p_a,\rho_a)$ must take. For example, consider a the scattering amplitude of three particles where particles $1$, $2$, and $3$ have spin-$s_1$, helicity-$h_2$ and helicity-$h_3$ respectively
. We would represent this object as,
\begin{align}
  M^{\{I_1,\dots,I_{2s_1}\},\{h_2\},\{h_3\}}(p_1,p_2,p_3) \,,
\end{align}
where the indices $\{I_1,\dots,I_{2s_1}\}$ are fully symmetrized. Under a little group transformation, we have,
\begin{align}\label{eq:amp:littlegroup:covariance}
  \c{A}^{\{I_1,\dots,I_{2s_1}\},\{h_2\},\{h_3\}} &= (W_{1L_1}^{I_1}\dots W_{1L_{2s_1}}^{I_{2s_1}})(w^{2h_2})(w^{2h_3})  M^{\{L_1,\dots,L_{2s_1}\},\{h_2\},\{h_3\}} \,,
\end{align}
where $W$ are $SU(2)$ transformations in the spin-$\frac{1}{2}$ representation and $w = e^{i\theta}$ are $U(1)$ transformations in the helicity-$\frac{1}{2}$ representation. 

\subsection{Helicity and Spin Spinors}
Given the Lorentz invariant and little group covariant structure of the amplitude, it would be great if we had variables that transformed under both the Lorentz group and little group. This is in fact the usefulness of `helicity-spinors' and `spin-spinors'. 
Helicity-spinors are introduced as objects that transform under both the $SL(2,C)$ Lorentz group and $U(1)$ little group, while spin-spinors are objects that transform under both the $SL(2,C)$ Lorentz group and $SU(2)$ little group. 

Thus, the amplitude can be written as a function of these variables,
\begin{align}\label{eq:amp:function_of_spinors}
    \c{A}^{\{I_1,\dots,I_{2s_1}\},\{h_2\},\{h_3\}}(p_1,p_2,p_3) &= \c{A}(\{\b{\lambda}_1^{\{I_1},\dots,\b{\lambda}_1^{I_{2s_1}\}}\},\{\lambda_2,\tilde\lambda_2,h_2\},\{\lambda_3,\tilde\lambda_3,h_3\})
\end{align}
where the helicity spinors $\lambda$ are unbolded and the spin spinors are bolded by convention (introduced in \cite{Arkani-Hamed:2017jhn}). Since the bolded spin spinors are always symmetrized, we can remove the explicit indices and infer the spin by keeping in mind that we need $2S$ spin spinors for a particle of spin $S$. Thus, we will often write instead,
\begin{align}
    \c{A}^{\{I_1,\dots,I_{2s_1}\},\{h_2\},\{h_3\}}(p_1,p_2,p_3) &= \c{A}(\{\overbrace{\b{\lambda}_1,\dots,\b{\lambda}_1}^{2S\,\text{times}} \},\{\lambda_2,\tilde\lambda_2,h_2\},\{\lambda_3,\tilde\lambda_3,h_3\}) \,,
\end{align}

In appendix \ref{sec:kinematics}, we demonstrate how to construct helicity and spin spinors, that have the correct transformation properties. Specifically, helicity spinors are defined to transform under the $h=\pm \frac{1}{2}$ representation of $U(1)$,
\begin{alignat}{4}
  \lambda_{\alpha} &\rightarrow e^{i2h\theta}\lambda_{\alpha} &&= e^{i\theta}\lambda_{\alpha} &&= w\lambda_{\alpha} \,, \nonumber \\
   \tilde\lambda_{\dot\alpha} &\rightarrow e^{-i2h\theta}\tilde\lambda_{\dot\alpha}&&=e^{-i\theta}\tilde\lambda_{\dot\alpha} &&=w^{-1}\tilde\lambda_{\dot\alpha}\,.
\end{alignat}
Spin-spinors are defined to transform under the $S=\frac{1}{2}$ representation, or a rank-$1$ (trivially) symmetric tensor,
\begin{align}
  \b{\lambda}_\alpha^{\;\;I} \rightarrow W_J^{\;\;I} \b{\lambda}_\alpha^{\;\;J} \qquad \text{and} \qquad \tilde{\b{\lambda}}_{\dot\alpha I} \rightarrow (W^{-1})_I^{\;\;J} \tilde{\b{\lambda}}_{\dot\alpha J} \,,
\end{align}
with $W$ an $SU(2)$ transformation in the spin-$\frac{1}{2}$ representation, where $I,J$ are the $SU(2)$ little group indices. The choice of $\b{\lambda}$ or $\tilde{\b{\lambda}}$ is irrelevant as one can use the Dirac equation (see Appendix \ref{sec:kinematics}) to freely convert from one to another, which is why in \eqref{eq:amp:function_of_spinors} we have only chosen one.

Transformations under the  Lorentz group are encoded in the $SL(2,C)$ indices $\alpha$ and $\dot\alpha$. With these spinors available, one can contract the Lorentz indices to form Lorentz invariant objects, that are still covariant with respect to the little group, which are exactly the type of objects that correspond to scattering amplitudes! 

For three-particle amplitudes, building these objects to have the proper little group covariance, is one of the central constraints, and a systematic analysis of all possible three particle amplitudes has been carried out in \cite{Arkani-Hamed:2017jhn}, and applied to the SM in \cite{Christensen:2018zcq}.

\section{Massless Three and Four Particle Amplitudes\label{sec:MasslessThreeandFourParticleAmplitudes}}

In this section, we review the construction of massless three and four particle amplitudes, and show how these naturally lead to the structures of Yang-Mills and Yukawa theories. We begin with the standard three particle amplitude to assist the reader with our conventions. Then, we construct tree level four particle amplitudes by gluing three particle amplitudes. By demanding unitary in the form of consistent factorization, we derive constraints on couplings which are identical to those one would attain from a gauge symmetry. We defer the reader to \cite{Elvang:2015rqa,Arkani-Hamed:2017jhn} for more details.

\subsection{Three Particle Amplitudes\label{sec:ThreeParticleAmplitudes}}

Massless three particle kinematics forces either $\lambda_1 \propto \lambda_2 \propto \lambda_3$ or $\tilde\lambda_1 \propto \tilde\lambda_2 \propto \tilde\lambda_3$. Lets consider the first case, so a massless three particle amplitude will only be a function of $\lambda_1,\lambda_2,\lambda_3$, which, to match the notation in eq.~\eqref{eq:amp:function_of_spinors},
\begin{align}
    \c{A}^{\{h_1\},\{h_2\},\{h_3\}}(p_1,p_2,p_3) = \c{A}(\{\lambda_1,h_1\},\{\lambda_2,h_2\},\{\lambda_3,h_3\})\,.
\end{align}
The most general Lorentz invariant object we can construct is given by making all possible contractions on the $SL(2,C)$ indices, 
\begin{align}
    \c{A}(\{\lambda_1,h_1\},\{\lambda_2,h_2\},\{\lambda_3,h_3\}) = \la12\ra^a\la23\ra^b\la31\ra^c\,,
\end{align}
where $a,b,c$ are unspecified. To fix $a,b,c$ we must impose that the amplitude has the correct little group covariance as in eq.~\eqref{eq:amp:littlegroup:covariance}. That is, under a little group transformation, we must have, $\c{A}^{\{h_1\},\{h_2\},\{h_3\}} \rightarrow (w_1^{2h_1})(w_2^{2h_2})(w_3^{2h_3})\c{A}^{\{h_1\},\{h_2\},\{h_3\}}$. Our anstaz, under a little group transformation, takes the form
\begin{align}
  \la12\ra^a\la23\ra^b\la31\ra^c\rightarrow w_1^{a+c}w_2^{a+b}w_3^{b+c}  \la12\ra^a\la23\ra^b\la31\ra^c\,.
\end{align}
Solving these constraints for $a,b,c$ yields, 
\begin{align}
\c{A}(\{\lambda_1,h_1\},\{\lambda_2,h_2\},\{\lambda_3,h_3\}) = \la12\ra^{h_1+h_2-h_3}\la23\ra^{h_2+h_3-h_1}\la31\ra^{h_3+h_1-h_2} \,.
\end{align}
An identical analysis can be performed assuming $\tilde\lambda_1 \propto \tilde\lambda_2 \propto \tilde\lambda_3$, and yields 
\begin{align}
   \c{A}(\{\tilde\lambda_1,h_1\},\{\tilde\lambda_2,h_2\},\{\tilde\lambda_3,h_3\}) = [12]^{-h_1-h_2+h_3}[23]^{-h_2-h_3+h_1}[31]^{-h_3-h_1+h_2} \,.
\end{align}
For a given set of helicities $h_1,h_2,h_3$, to distinguish which of these to use, we must impose locality, which constrains the mass dimension of the momentum dependence (a three particle amplitude in four dimensions must have mass dimension one). This yields
\begin{alignat}{3}
\label{eq:3particlemassless}
    \c{A}^{\{h_1\},\{h_2\},\{h_3\}} &= g\, \langle 12\rangle^{h_1+h_2-h_3} \langle 23\rangle^{h_2+h_3-h_1} \langle 31\rangle^{h_3+h_1-h_2} \,,\quad &\text{ if}\quad h_1+h_2+h_2 >0 \:\: \nonumber\\
 & =\tilde{g}\, [12]^{h_3-h_1-h_2} [23]^{h_1-h_2-h_3} [31]^{h_2-h_3-h_1}\,,\qquad &\text{ if}\quad h_1+h_2+h_2 < 0 \: .
\end{alignat} 

Now that we have fixed the kinematic form of the amplitude, we can return to the ignored $\sigma_a$ labels. As mentioned before, these encode any additional quantum numbers that the particle can take. The only contractions we have left are of the form $_{\text{out}}\la\sigma_1,\dots,\sigma_n|0\ra_{\text{in}}$, which, in the simplest sense, say that the sum of `charges' is zero, assuming all outgoing. For a three particle vertex, between possibly different particles with labels $\sigma_a$, $\sigma_b$ and $\sigma_c$, we attach a simple function $f(\sigma_a,\sigma_b,\sigma_c)$ to the vertex. In the next section, we will see how the properties of these functions can be connected to kinematics.

\subsection{Four Particle Amplitudes}

Let us move on to four particle amplitudes. The additional constraint on the structure of the amplitude comes from unitarity in form of consistent factorization. That is, when some internal momentum goes on-shell, the residue must be factorizable as a product of lower point amplitudes. In this section, we only consider massless particles, so the condition takes the form,
\begin{align}
    \c{A} &= \frac{\c{A}_L^{a\; h} \c{A}_R^{a\; -h}}{P^2}\,,
\end{align}
where $a$ is the index of some intermediate particle.

It is by now well known that consistent factorization on the amplitude $\c{A}(1_i^{+s},2_a^{+1},3_b^{-1},4_j^{-s})$ leads to the discovery of the familiar Yang-Mills structure of spin $0$, $1/2$ and $1$ particles \cite{Arkani-Hamed:2017jhn,Benincasa:2007xk}. We follow similar logic, and demand consistent factorization on the amplitude $\c{A}(1_j^{+s},2_I^{0},3_a^{-1},4_k^{-s})$. In doing so, we discover the familiar structures of Yukawa theories. These structures are then used in Section \ref{sec:MatchingSolutions} to show the Higgs mechanism is necessary for UV-IR consistency.

First, we will present the analysis in \cite{Arkani-Hamed:2017jhn} for the amplitude $\c{A}(1_i^{+s},2_a^{+1},3_b^{-1},4_j^{-s})$ to set notation and conventions. Then, we will repeat the analysis on the amplitude $\c{A}(1_j^{+1/2},2_I^{0},3_a^{-1},4_k^{-1/2})$ to derive similarly interesting constraints.

We begin by noting that the amplitude $\c{A}(1_i^{+s},2_a^{+1},3_b^{-1},4_j^{-s})$ has three channels given by,
\begin{align}
    \vcenter{\hbox{\includegraphics[scale=.8]{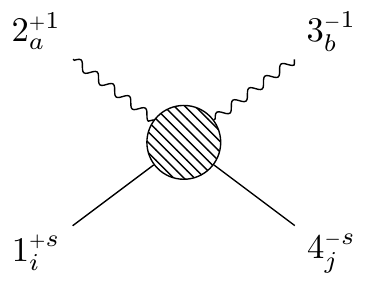}}} &\supset\vcenter{\hbox{\includegraphics[scale=.8]{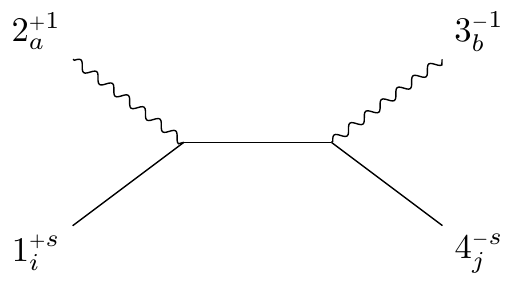}}} + \vcenter{\hbox{\includegraphics[scale=.8]{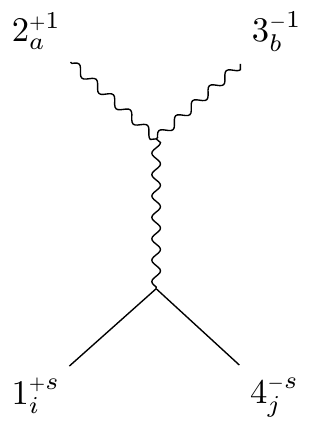}}} + \vcenter{\hbox{\includegraphics[scale=.8]{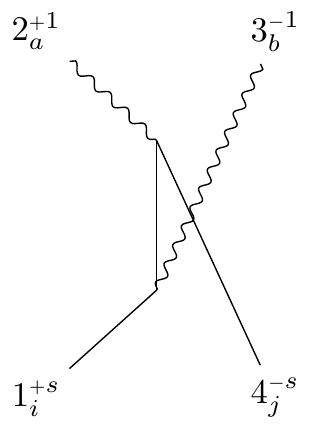}}} \label{eq:Compton_all}   \,.
\end{align}
The residues in every channel can be obtained by gluing $M_L$ and $M_R$. When we have several spin-$s$ particles $i,j$ coupling to a spin-$1$ particle $a$, we attach the vertex $(T^R_a)_{ij}$. Note that $R=s$ for the representation the scalars ($s=0$) transform under, $R=f$ for the representation the fermions ($s=1/2$) transform under, and lastly $R = \text{ad}$ for the (adjoint) representation the bosons ($s=1$) transform under. The residues in every channel takes the form $(\la12\ra[34])^{2s}\la2|1|3]^{2-2s} \times r$ where,
\begin{align}\label{eq:compton:residues}
    r_s &= (T^R_aT^R_b)_{ij}\times\frac{1}{u} \,,\nonumber \\
    r_u &= (T^R_bT^R_a)_{ij}\times\frac{1}{s}\,,\nonumber \\
    r_t &= (T^R_c)_{ij}(T^{\text{Ad}}_c)_{ab}\times\frac{1}{2}\left(\frac{1}{s} - \frac{1}{u}\right) \,.
\end{align}
Given the structure of the residues, for $s=\{0,1/2,1\}$, the ansatz for the four particle amplitude has the form,
\begin{align}\label{eq:compton:ansatz}
   \c{A}(1_i^{+s},2_a^{+1},3_b^{-1},4_j^{-s}) =  (\la12\ra[34])^{2s}\la2|1|3]^{2-2s}\left( \frac{A_{ijab}}{su} +  \frac{B_{ijab}}{ut} +  \frac{C_{ijab}}{ts}\right) \,.
\end{align}
Matching the residues in the ansatz above to those computed by gluing in eq.~\eqref{eq:compton:residues} yields,
\begin{alignat}{3}
    r_s &= (T^R_aT^R_b)_{ij}\times\frac{1}{u}&&= \left(A_{ijab} - C_{ijab} \right)\frac{1}{u} \,, \nonumber \\ 
    r_u &= (T^R_bT^R_a)_{ij}\times\frac{1}{s}&&= \left(A_{ijab} - B_{ijab} \right)\frac{1}{s} \,,\nonumber \\ 
    r_t &= (T^R_c)_{ij}(T^{\text{Ad}}_c)_{ab}\times\frac{1}{s}&&= \left(-B_{ijab} + C_{ijab} \right)\frac{1}{s} \,,
\end{alignat}
which has solutions only if $[T_b^R,T_a^R]_{ij} = (T^R_c)_{ij}(T^{\text{Ad}}_c)_{ab}$. Identifying $(T^{\text{Ad}}_c)_{ab} = -f_{abc}$ yields the familiar
\begin{align}\label{eq:yangmillstructure}
    [T^R_a,T^R_b]_{ij} &= f_{ab}^{\;\;\;c}\;(T^R_c)_{ij} \,.
\end{align}

When we consider the case of $s=1$, then $i,j = d,e$, and $R = \text{Ad}$. Along with $(T^{\text{Ad}}_c)_{ab} = -f_{abc}$, we discover that the coefficients $f_{abc}$ must satisfy the Jacobi identity,
\begin{align}\label{eq:jacobi}
     f_{ade}f_{ebc} + f_{bde}f_{eac} + f_{abe}f_{ecd} = 0 \,.
\end{align}
We defer to \cite{Arkani-Hamed:2017jhn} for further exploration and discussion on these amplitudes.

Next, we move on to the amplitude $\c{A}(1_j^{-1/2},2_I^{0},3_a^{+1},4_k^{-1/2})$. By demanding consistent factorization identical to the above, we will derive constraints between the couplings and the matrices $T_a^R$ that are identical to those one would obtain by assuming the existence of a gauge symmetry. The amplitude $\c{A}(1_j^{-1/2},2_I^{0},3_a^{+1},4_k^{-1/2})$ has three channels given by
\begin{align}
    \vcenter{\hbox{\includegraphics[scale=.8]{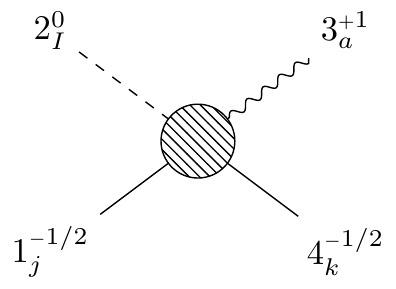}}} &\supset\vcenter{\hbox{\includegraphics[scale=.8]{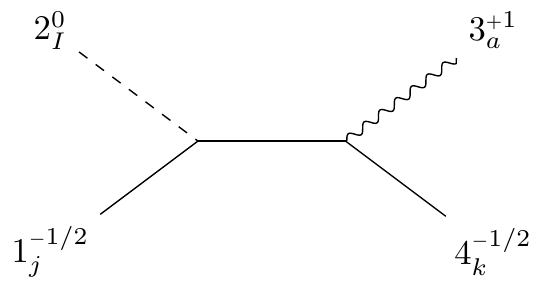}}} + \vcenter{\hbox{\includegraphics[scale=.8]{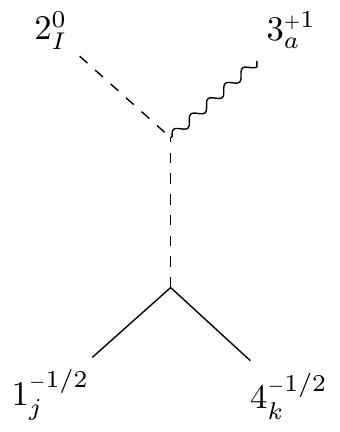}}} + \vcenter{\hbox{\includegraphics[scale=.8]{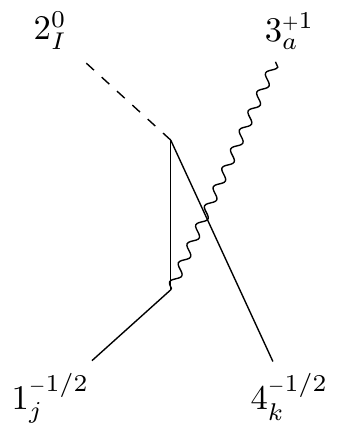}}} \label{eq:Yukawa_all} \,.
\end{align}
When we have several spin-$1/2$ particles $i,j$ coupling to a spin-$0$ particle $I$, we attach the vertex $(Y_I)_{ij}$. The residues in every channel take the form $\la13\ra\la34\ra[14]^2 \times r$ where,
\begin{align}\label{eq:yukawa:residues}
    r_s &= +(Y_I)_{jl}(T_a^f)_{lk} \times \frac{1}{t} \,, \nonumber \\
    r_u &= -({Y}_I)_{kl}(T_a^f)_{lj} \times \frac{1}{t} \,,\nonumber \\
    r_t &= +({Y}_L)_{jk}(T_a^s)_{LI} \times \frac{1}{u} \,.
\end{align}
Given the structure of the residues, the ansatz for the four particle amplitude has the form,
\begin{align}\label{eq:yukawa:ansatz}
   \c{A}(1_j^{+1/2},2_I^{0},3_a^{-1},4_k^{-1/2}) =  \la13\ra\la34\ra[14]^2 \left( \frac{A_{ijab}}{su} +  \frac{B_{ijab}}{ut} +  \frac{C_{ijab}}{ts}\right) \,.
\end{align}
Again, matching the residues in the ansatz, to those computed by gluing in eq.~\eqref{eq:yukawa:ansatz} yields the condition,
\begin{alignat}{3}
    r_s &= +(Y_I)_{jl}(T_a^f)_{lk} \times \frac{1}{t}&&= \left(-A_{jIak} + C_{jIak} \right)\frac{1}{t} \,, \nonumber \\ 
    r_u &= -({Y}_I)_{kl}(T_a^f)_{lj} \times \frac{1}{t}&&= \left(-A_{jIak} +B_{jIak} \right)\frac{1}{t} \,, \nonumber \\ 
    r_t &= +({Y}_L)_{jk}(T_a^s)_{LI} \times \frac{1}{u} &&= \left(B_{jIak} - C_{jIak} \right)\frac{1}{u} \,.\,
\end{alignat}
which has solutions only if,
\begin{align}\label{eq:yukawa:constraint}
    (Y_I)_{jl}(T_a^f)_{lk} + ({Y}_I)_{lk}(T_a^f)_{lj} + ({Y}_L)_{jk}(T_a^s)_{LI}= 0 \,.
\end{align}
This is exactly the constraint we would get assuming a gauge theory. To illustrate, we know ${Y}_{Ijk}$ picks out a singlet as ${Y}_{Ijk}\phi^I\psi^j\psi^k$. Applying the gauge transformations $\phi^{I^\prime} = (e^{i T^s_a})^{I^\prime}_{\;\;I}\phi^I$ and $\psi^{I^\prime} = (e^{i T^f_a})^{i^\prime}_{\;\;i}\psi^i$ to the transformation
\begin{align}
    {Y}_{Ijk}\phi^I\psi^j\psi^k &\rightarrow {Y}_{I^\prime j^\prime k^\prime }\phi^{I^\prime}\psi^{j^\prime}\psi^{k^\prime} \nonumber \\
    &= {Y}^f_{I^\prime j^\prime k^\prime }\;(e^{i T^s_a})^{I^\prime}_{\;\;I}\phi^I \;(e^{i T^f_a})^{j^\prime}_{\;\;j}\psi^j \;(e^{i T^f_a})^{k^\prime}_{\;\;k}\psi^k \,,
\end{align}
at first order yields eq.~\eqref{eq:yukawa:constraint}.

Lastly, lets consider the special case of eq.~\eqref{eq:yukawa:constraint}, where $({Y}_L)_{jk}(T_a^s)_{LI} = 0$. Since $(T_a^s)_{LI}$ only acts on the $L$ scalar degrees of freedom, we can separate its action as $({Y}_L)_{jk} = n_L \tilde{m}_{jk}$. Eq.~\eqref{eq:yukawa:constraint} now reads,
\begin{align}\label{eq:yukawa:constraint:gauge:mass}
    \tilde{m}_{jl}(T_a^f)_{lk} + \tilde{m}_{lk}(T_a^f)_{lj} = 0 \,,
\end{align}
which is the condition needed for background scalars to contribute to fermion masses. 
We can see this by applying the gauge transformations $\psi^{I^\prime} = (e^{i T^f_a})^{i^\prime}_{\;\;i}\psi^i$ to
\begin{align}
    {\tilde{m}}_{jk}\psi^j\psi^k &\rightarrow {\tilde{m}}_{j^\prime k^\prime}\psi^{j^\prime}\psi^{k^\prime} \nonumber \\
    &= {\tilde{m}}_{j^\prime k^\prime } \;(e^{i T^f_a})^{j^\prime}_{\;\;j}\psi^j \;(e^{i T^f_a})^{k^\prime}_{\;\;k}\psi^k \,,
\end{align}
at first order yields eq.~\eqref{eq:yukawa:constraint:gauge:mass}. 

\section{\texorpdfstring{Non-Abelian Higgs: $G \rightarrow H$}{Non-Abelian Higgs G to H}\label{sec:Non-AbelianHiggs}}

Much of the simplicity will be gained by considering the action of the full symmetry group $G$ at once, rather than working with its subgroups $G_1, \dots, G_n$. In the real world, and many toy examples, we usually specify the subgroups and assign different coupling constants to them. For example, in the Standard Model, the internal symmetry group is $G = SU(3) \times SU(2) \times U(1)$, to which we can assign a strong, weak and hypercharge coupling constant. Although this distinction is useful, we find that by reliving ourselves of the subgroups, in the appropriate manner, can elucidate additional structures in these amplitudes.

The strategy we will take is to define a UV and IR theory, through the three particle scattering amplitudes, and demand that the high energy limit of the IR theory can match onto the UV theory. In doing this, we will rediscover many familiar results about spontaneous symmetry breaking.

We will use hatted indexes in the UV and unhatted indexes in the IR. Since the spinors for massive particle momenta are bolded, and massless are unbolded, we will bold any IR particles in our diagrams for visual consistency and aid.

\subsection{The UV}

In the UV, all the particle labels are hatted. We have a symmetry group $G$ of dimension $d_G$.  In general, $G$ can be expressed as a direct product of simple compact Lie groups and $U(1)$,
\begin{align}
    G &=  G_1 \times \dots \times  G_i \times \dots \times  G_n \,.
\end{align}
The associated Lie algebra $\mathfrak{\tilde g}$ (we will drop all the tildes soon) is spanned by $d_G$ generators 
\begin{equation}
  \mathfrak{\tilde g}= \{\tilde T_1,\dots,\tilde {T}_{\hat{a}} ,\dots,\tilde T_{d_G}\} \,.
\end{equation}
There is a separate coupling $g_a$ for each simple group or $U(1)$ factor of the group $G$. The generators, $\tilde{T}_{\hat{a}}$, thus separate out into different classes, and $g_{\hat{a}} = g_{\hat{b}}$ if $\tilde{T}_{\hat{a}}$ and $\tilde{T}_{\hat{b}}$ are in the same class. Of course, by standard definitions, $[\tilde{T}_{\hat{a}},\tilde{T}_{\hat{b}}] = \tilde{f}_{\hat{a}\hat{b}\hat{c}} \tilde{T}_{\hat{c}}$. As mentioned earlier, we find it convenient to re-scale the basis of our Lie algebra as $T_{\hat{a}} = g_{\hat{a}} \tilde T_{\hat{a}}$ (no sum over $\hat{a}$, and tildes now removed),
\begin{align}
   \mathfrak{g} &= \left\{ \{g_i\tilde T_{1_i}, g_i\tilde T_{2_i}, \dots g_i\tilde T_{n_i}\},\{ g_j\tilde T_{1_j},  g_j\tilde T_{2_j}, \dots ,  g_j\tilde T_{n_j}\}, \dots ,\{ g_n\tilde T_{1_n}\} \right\} \,, \nn \\
   &= \{  T_1,  T_2, \dots , T_{d_G}\}\,.
\end{align}
Note that this statement is representation independent. These (rescaled) generators will obey the commutation relation in eq.~\eqref{eq:yangmillstructure}, with hatted indices,
\begin{align}
   [ T_{\hat{a}}, T_{\hat{b}}] =   f_{\hat{a}\hat{b}}^{\;\;\; \hat{c}}  T_{\hat{c}} \,,
\end{align}
where $f_{\hat{a}\hat{b}\hat{c}} = \frac{g_{\hat{a}}g_{\hat{b}}}{g_{\hat{c}}} \tilde{f}_{\hat{a}\hat{b}\hat{c}}$. Given a general set of generators and couplings, one can easily compute the new structure constants via,
\begin{align}\label{eq:compute:structureconstant}
    f_{\hat{a}\hat{b}\hat{c}} = \frac{1}{\text{Tr}\{T_{\hat{a}} T_{\hat{a}}\}}\text{Tr}\{[T_{\hat{a}},T_{\hat{b}}]T_{\hat{c}}\} \,.
\end{align}

The spectrum consist of $N_s$ spin-$0$, $N_f$ spin-$1/2$ and $N_{\text{ad}}$ spin-$1$ particles, which transform under (possibly reducible) representations $R=s$, $R=f$ and $R={\text{ad}}$ of the symmetry group $G$ respectively. So, the number of particles is equal to the dimension of the representation, that is, $N_s = \text{dim}(R=s)$, $N_f = \text{dim}(R=f)$, and $N_{\text{ad}} = \text{dim}(R={\text{ad}}) = d_G$. The scalars are labelled by $\{\hat{I},\hat{J},\hat{K}\}$, the fermions by $\{\hat{i},\hat{j},\hat{k}\}$, and the bosons by $\{\hat{a},\hat{b},\hat{c}\}$. We summarize the notation in Table~\ref{table:UV_notation}.

We consider all particles of the same helicity to be identical, and assume all our particles are real. This may seem like a over simplicity, however, since we are working with (possibly) reducible representations, this will not be an obstruction in considering more complicated models. Hence, we allow $SO(\text{dim}(R))$ symmetries within the same helicity. To elaborate, under free propagation, the spin-$0$, spin-$1/2$ and spin-$1$ particles have an $SO(N_s)$, $SO(N_f)$ and $SO(d_G)$ symmetry that we will represent with $U$, $\Omega$ and $\c{O}$ respectively.  We will exploit these in the matching process.

\begin{table}[h!]
\centering
 \begin{tabular}{|c | c | c | c | c |} 
 \hline
 Spin & Rep., $R$ & $\text{dim}(R)$ & Labels & $SO(\text{dim}(R))$\\
 \hline
 $0$ & $s$ & $N_s$ & $\hat{I},\hat{J},\hat{K}$ & $U$ \\ 
 $\frac{1}{2}$ & $f$ & $N_f$ & $\hat{i},\hat{j},\hat{k}$ & $\Omega$\\
 $1$ & $\text{ad}$ & $N_{\text{ad}}$ & $\hat{a},\hat{b},\hat{c}$ & $\c{O}$\\
 \hline
\end{tabular}
\caption{Summary of UV spectrum.}
\label{table:UV_notation}
\end{table}

Now that we have defined the spectrum and symmetry group, we can move on to interactions. We will consider only the 3-particle amplitudes. To each of these amplitudes, we assign a coupling. These couplings must obey the constraints from consistent factorization, notably eqns.~\eqref{eq:yangmillstructure}, \eqref{eq:jacobi} and \eqref{eq:yukawa:constraint}. We only consider amplitudes with $\sum h = \pm 1$, as these correspond to amplitudes that have dimensionless coupling constants (see \cite{McGady:2013sga} for general considerations). We list the amplitudes considered and elaborate on any properties that will be useful:
\begin{itemize}
  \item Three spin one particles:

  \begin{align}\label{eq:uvamp:1}
     \vcenter{\hbox{\includegraphics{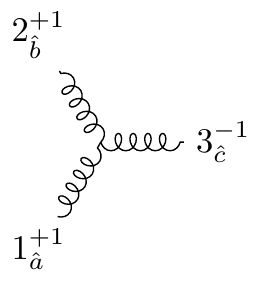}}}    &= \c{A}(1^{+1}_{\hat{a}},2^{+1}_{\hat{b}},3^{-1}_{\hat{c}}) = f_{\hat{a}\hat{b}{\hat{c}}} \; \c{A}(1^{+1},2^{+1},3^{-1})  \,,
  \end{align}
  where $ \c{A}(1^{+1},2^{+1},3^{-1}) = \frac{\la 12 \ra^3}{\la 23\ra \la 31 \ra}$. Note that, since particles $1$ and $2$ have the same helicity, this amplitude should be even under the exchange of the bosons $1^{+1}_{\hat{a}} \leftrightarrow 2^{+1}_{\hat{b}}$, so that $\c{A}(1^{+1}_{\hat{a}},2^{+1}_{\hat{b}},3^{-1}_{\hat{c}}) = \c{A}(2^{+1}_{\hat{b}},1^{+1}_{\hat{a}},3^{-1}_{\hat{c}})$. Since $\c{A}(1^{+1},2^{+1},3^{-1}) = - \c{A}(2^{+1},1^{+1},3^{-1})$, then we must have the antisymmetric property $f_{\hat{a}\hat{b}{\hat{c}}} = - f_{\hat{b}\hat{a}{\hat{c}}}$.

\item Two spin zero and one spin one:
\begin{align}\label{eq:uvamp:2}
   \vcenter{\hbox{\includegraphics{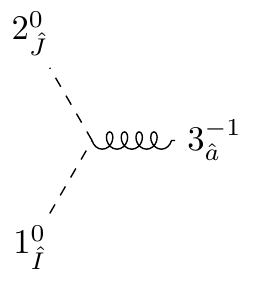}}}    &= \c{A}(1^{0}_{\hat{I}},2^{0}_{\hat{J}},3^{-1}_{\hat{a}}) = (T^s_{\hat{a}})_{\hat{I}\hat{J}} \; \c{A}(1^{0},2^{0},3^{-1})  \,,
\end{align}
with $\c{A}(1^{0},2^{0},3^{-1}) = \frac{[23][31]}{[12]}$. By demanding that the amplitude is even under the exchange of the bosons $1_{\hat{I}}^0 \leftrightarrow 2_{\hat{J}}^0$, and following the same logic as previously, we find the antisymmetric property $(T^s_{\hat{a}})_{\hat{I}\hat{J}} = - (T^s_{\hat{a}})_{\hat{J}\hat{I}}$.

\item Two spin half and one spin one:
\begin{align}\label{eq:uvamp:3}
   \vcenter{\hbox{\includegraphics{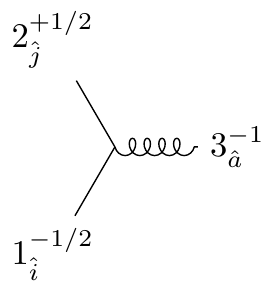}}}    &= \c{A}(1^{-1/2}_{\hat{i}},2^{+1/2}_{\hat{j}},3^{+1}_{\hat{a}}) = (T^f_{\hat{a}})_{\hat{i}\hat{j}} \; \c{A}(1^{-1/2},2^{+1/2},3^{-1})  \,,
\end{align}
with $ \c{A}(1^{-1/2},2^{+1/2},3^{-1}) = \frac{[31]^2}{[12]}$. In this case, all the particles are distinguishable, and so we have no further constraints on $(T^f_{\hat{a}})_{\hat{j}\hat{i}}$.

\item Two spin half and one spin zero:
\begin{align}\label{eq:uvamp:4}
   \vcenter{\hbox{\includegraphics{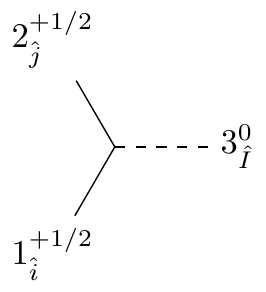}}}    &= \c{A}(1^{+1/2}_{\hat{i}},2^{+1/2}_{\hat{j}},3^{0}_{\hat{I}}) = Y^s_{\hat{I}\hat{i}\hat{j}} \; \c{A}(1^{+1/2},2^{+1/2},3^{0})  \,,
\end{align}
where $\c{A}(1^{+1/2},2^{+1/2},3^{0}) = \la 12 \ra$. Demanding the amplitude is odd under the exchange of the fermions $1^{+1/2}_{\hat{i}} \leftrightarrow 2^{+1/2}_{\hat{j}}$, tells us that the Yukawa coupling is symmetric in $Y_{\hat{I}\hat{i}\hat{j}} \leftrightarrow Y^s_{\hat{I}\hat{j}\hat{i}}$.
\end{itemize}

The four amplitudes above define our UV theory. Next, we will move on to define our IR theory.

\subsection{The IR}

In the IR, all particle labels are unhatted. There is a symmetry group, which the particles must respect when interacting, but instead of imposing a group in advance, we will try to discover what it is allowed, if it is to be consistent with the UV theory defined previously.

The spectrum consist of massless and massive particles. Again, we will only consider spins $0$, $1/2$ and $1$. Since the spectrum is slightly more complicated, we will continue to use the un-bolded and bolded notation for massless and massive particles respectively. For spin-$0$ particles, the massless particles are labeled by $\{I,J,K\}$ and the massive particles by $\{\r{I},\r{J},\r{K}\}$. For spin-$1/2$ particles, the massless particles are labeled by $\{i,j,k\}$ and the massive by $\{\r{i},\r{j},\r{k}\}$. Lastly, for spin-$1$ particles, the massless particles are labelled by $\{a,b,c\}$ and the massive by $\{\r{a},\r{b},\r{c}\}$. Keep in mind, all labels are unhatted. We summarize the labels in Table~\ref{table:IR_notation}.

\begin{table}[h!]
\centering
 \begin{tabular}{|c | c | c | c |} 
 \hline
 Spin & Massless Labels & Massive Labels & Collectively \\
 \hline
 $0$ & $I,J,K$ & $\r{I},\r{J},\r{K}$  &  $\u{I},\u{J},\u{K}$\\ 
 $\frac{1}{2}$ & $i,j,k$ &  $\r{i},\r{j},\r{k}$ & $\u{i},\u{j},\u{k}$ \\
 $1$ & $a,b,c$ & $\r{a},\r{b},\r{c}$ & $\u{a},\u{b},\u{c}$\\
 \hline
\end{tabular}
\caption{Summary of IR spectrum.}
\label{table:IR_notation}
\end{table}

Now that we have defined the spectrum, we can move onto the interactions. The presence of massless and massive particles in the IR makes constructing an exhaustive list quite cumbersome for our purposes.
We will instead take the approach of defining only the amplitudes with all massive external legs. Then, any other combination of massless and massive particles in the IR can be determined by taking the appropriate high energy limit of that leg, and un-bolding the particle label. For a more exhaustive list of these combinations amplitudes, see \cite{Christensen:2018zcq,Bachu:2019ehv,Arkani-Hamed:2017jhn}.

The relevant amplitudes we consider are:
\begin{itemize}

  \item Three massive spin one particles:
  \begin{align}\label{eq:Threemassivespioneparticles}
   \vcenter{\hbox{\includegraphics{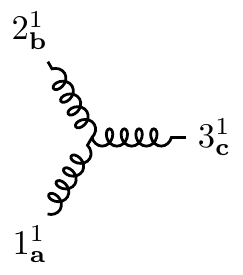}}}    &= \c{A}(\b{1}^{1}_{\r{a}},\b{2}^{1}_{\r{b}},\b{3}^{1}_{\r{c}}) = h_{\r{a}\r{b}\r{c}} \, \c{A}(\b{1}^{1},\b{2}^{1},\b{3}^{1})  \,,
  \end{align}
  where $ \c{A}(\b{1}^{1},\b{2}^{1},\b{3}^{1})  = \frac{1}{m_{\b{a}}m_{\b{b}}m_{\b{c}}}(\la\b{12}\ra[\b{12}]\la\b{3}|p_1-p_2|\b{3}] + \text{cyc.})$.

  \item Two massive spin half and one massive spin one:
  \begin{align}\label{eq:Two:massive:spin:half:and:one:massive:spin:one}
     \vcenter{\hbox{\includegraphics{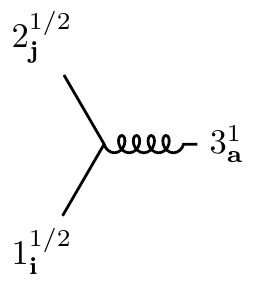}}}    = \c{A}(\b{1}^{1/2}_{\r{i}},\b{2}^{1/2}_{\r{j}},\b{3}^{1}_{\r{c}}) & = (X^{f}_{1\;\r{a}})_{\r{i}\r{j}} \la\b{13}\ra[\b{32}] + (X^{f}_{2\;\r{a}})_{\r{i}\r{j}}[\b{13}]\la\b{32}\ra \nonumber \\
                                    &+ (X^{f}_{3\;\r{a}})_{\r{i}\r{j}} \la\b{13}\ra\la\b{32}\ra + (X^{f}_{4\;\r{a}})_{\r{i}\r{j}} [\b{13}][\b{32}]\,.
  \end{align}

\end{itemize}

Note that in both eq.~\eqref{eq:Threemassivespioneparticles} and eq.~\eqref{eq:Two:massive:spin:half:and:one:massive:spin:one}, we have not specified any properties on the coefficients. We expect that they should conserve charge for some symmetry group, but we will discover their properties in the next section. 

Before we move on, we illustrate how the notation adapts to massless particles in the IR. For example, suppose we wanted to consider one massless helicity one particle $a$, and two massive spin one particles $\r{b}$ and $\r{c}$. The amplitude, given by $\c{A}(1_a^{+1},\b{2}_{\r{b}},\b{3}_{\r{c}})$, can be determined by taking the high energy limit of particle $\b{1}_a^1$ of $ \c{A}(\b{1}^{1}_{\r{a}},\b{2}^{1}_{\r{b}},\b{3}^{1}_{\r{c}})$, i.e. 
\begin{align}
     \c{A}(\b{1}^{1}_{\r{a}},\b{2}^{1}_{\r{b}},\b{3}^{1}_{\r{c}}) \xrightarrow{\text{HE limit of }1_{\r{a}}} \c{A}(1_a^{+1},\b{2}^1_{\r{b}},\b{3}^1_{\r{c}}) \,,
\end{align}
where $\c{A}(1_a^{+1},\b{2}^1_{\r{b}},\b{3}^1_{\r{c}}) \propto h_{a\r{b}\r{c}}$.
Although we have taken the high energy limit, we have remained in the IR, as our goal was simply to obtain the form of an amplitude with a massless leg in the IR. In the next section, we will consider taking high energy limits to map us onto the UV.

\subsection{UV IR Matching}

Now, we demand that the high energy limit of amplitudes in the IR match onto some combination of amplitudes in the UV. 

The process of matching can be summarized as:
\begin{enumerate}
    \item Select an amplitude in the IR.
    \item Project some spin configuration, and take the high energy limit of all particles.
    \item Demand that it is equivalent to some linear combination of amplitudes in the UV.
\end{enumerate}

For Step 1, our menu of IR amplitudes are eq.~\eqref{eq:Threemassivespioneparticles}, eq.~\eqref{eq:Two:massive:spin:half:and:one:massive:spin:one} and any variation with massless legs (which we mentioned how to construct). For Step 2, we will defer explicit calculations to Appendix~\ref{sec:HighEnergyLimits}, and point the reader to additional calculations in \cite{Christensen:2018zcq,Bachu:2019ehv,Arkani-Hamed:2017jhn}. For Step 3, our menu of UV amplitudes comprise of eq.~\eqref{eq:uvamp:1}, eq.~\eqref{eq:uvamp:2}, eq.~\eqref{eq:uvamp:3} and eq.~\eqref{eq:uvamp:4}, and we will use the $SO(\text{dim}(R))$ symmetries to construct the linear combinations.

We summarize the relevant results of this process below, and refer the reader to Tables~\ref{table:UV_notation} and \ref{table:IR_notation} for the index notation. 
\begin{itemize}
    \item $\c{A}(\b{1}^{+1}_\r{a},\b{2}^{+1}_\r{b},\b{1}^{-1}_\r{c})$
    \begin{align}\label{eq:match:boson:trans}
     \vcenter{\hbox{\includegraphics[scale=1]{figs/IR_gluons_abc.pdf}}} \overset{HE} \longrightarrow \vcenter{\hbox{\includegraphics[scale=1]{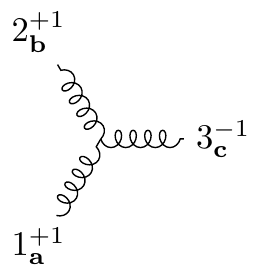}}}  & \equiv \c{O}^{\hat{a}}_{\;\; \r{a}} \c{O}^{\hat{b}}_{\;\; \r{b}} \c{O}^{\hat{c}}_{\;\; \r{c}} \left(\vcenter{\hbox{\includegraphics[scale=1]{figs/UV_gluons_abc.pdf}}}\right) \nn \\
     \fir_{\r{a}\r{b}\r{c}}\c{A}(1^{+1},2^{+1},3^{-1}) & \equiv \c{O}^{\hat{a}}_{\;\; \r{a}} \c{O}^{\hat{b}}_{\;\; \r{b}} \c{O}^{\hat{c}}_{\;\; \r{c}}\;\fuv_{\hat{a}\hat{b}\hat{c}} \c{A}(1^{+1},2^{+1},3^{-1}) 
    \end{align}

    \item $\c{A}(\b{1}^{0}_\r{a},\b{2}^{0}_\r{b},\b{1}^{-1}_\r{c})$
    \begin{align}\label{eq:match:boson:long}
         \vcenter{\hbox{\includegraphics[scale=1]{figs/IR_gluons_abc.pdf}}} \overset{HE} \longrightarrow \vcenter{\hbox{\includegraphics[scale=1]{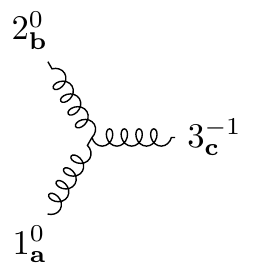}}}  & \equiv \c{U}^{\hat{I}}_{\;\; \r{a}}\;\c{U}^{\hat{J}}_{\;\; \r{b}}\;\c{O}^{\hat{a}}_{\;\; \r{c}} \left(\vcenter{\hbox{\includegraphics[scale=1]{figs/UV_scalarscalargluon.pdf}}}\right) \nn \\
         \fir_{\r{a}\r{b}\r{c}}\frac{(m_\r{c}^2 - m_\r{a}^2 -m_\r{b}^2)}{2\,m_\r{a}m_\r{b}}\c{A}(1^{0},2^{0},3^{-1}) & \equiv \c{U}^{\hat{I}}_{\;\; \r{a}}\;\c{U}^{\hat{J}}_{\;\; \r{b}}\;\c{O}^{\hat{a}}_{\;\; \r{c}}\;(T^s_{\hat{a}})_{\hat{I}\hat{J}} \c{A}(1^{0},2^{0},3^{-1})
    \end{align}

    \item $\c{A}(\b{1}^{+1/2}_\r{i},\b{2}^{+1/2}_\r{j},\b{1}^{+1}_\r{a})$
    \begin{align}\label{eq:match:fermion:allplus}
         \vcenter{\hbox{\includegraphics[scale=1]{figs/IR_fermionfermiongluon.pdf}}} \overset{HE} \longrightarrow \vcenter{\hbox{\includegraphics[scale=1]{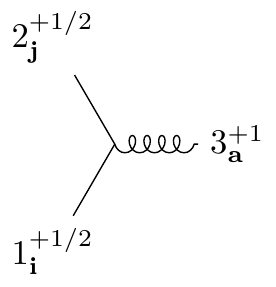}}}  & \equiv  0 \nn \\
        (X^{f}_{3\;\r{a}})_{\r{i}\r{j}}\c{A}(1^{+1/2},2^{+1/2},3^{+1}) & \equiv  0_{\r{a}\r{i}\r{j}}
    \end{align}

    \item $\c{A}(\b{1}^{-1/2}_\r{i},\b{2}^{-1/2}_\r{j},\b{1}^{-1}_\r{a})$
    \begin{align}\label{eq:match:fermion:allminus}
         \vcenter{\hbox{\includegraphics[scale=1]{figs/IR_fermionfermiongluon.pdf}}} \overset{HE} \longrightarrow \vcenter{\hbox{\includegraphics[scale=1]{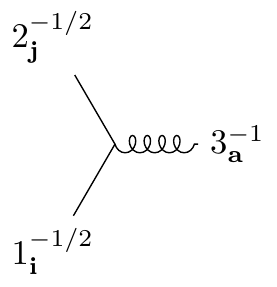}}}  & \equiv  0 \nn \\
        (X^{f}_{4\;\r{a}})_{\r{i}\r{j}}\c{A}(1^{-1/2},2^{-1/2},3^{-1}) & \equiv  0_{\r{a}\r{i}\r{j}}
    \end{align}

    \item $\c{A}(\b{1}^{-1/2}_\r{i},\b{2}^{+1/2}_\r{j},\b{1}^{+1}_\r{a})$
    \begin{align}\label{eq:match:fermion:trans1}
         \vcenter{\hbox{\includegraphics[scale=1]{figs/IR_fermionfermiongluon.pdf}}} \overset{HE} \longrightarrow \vcenter{\hbox{\includegraphics[scale=1]{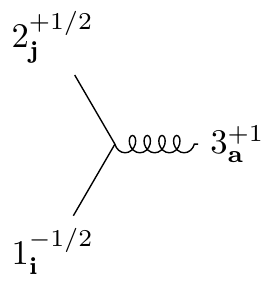}}}  & \equiv  \Omega^{\hat{i}}_{\;\;\r{i}} \; \Omega^{\hat{j}}_{\;\;\r{j}} \; \c{O}^{\hat{a}}_{\;\; \r{a}}  \left(\vcenter{\hbox{\includegraphics[scale=1]{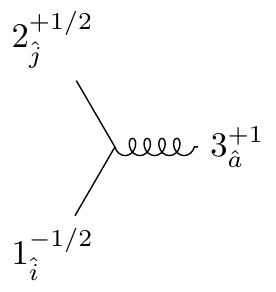}}}\right) \nn \\
       m_\r{a} (X^{f}_{2\;\r{a}} +X^{f}_{3\;\r{a}})_{\r{i}\r{j}}\c{A}(1^{-1/2},2^{+1/2},3^{+1}) & \equiv  \Omega^{\hat{i}}_{\;\;\r{i}} \; \Omega^{\hat{j}}_{\;\;\r{j}} \; \c{O}^{\hat{a}}_{\;\; a} (T^f_{\hat{a}})_{\hat{i}\hat{j}}  \c{A}(1^{-1/2},2^{+1/2},3^{+1}) 
    \end{align}

    \item $\c{A}(\b{1}^{+1/2}_\r{i},\b{2}^{-1/2}_\r{j},\b{1}^{+1}_\r{a})$
    \begin{align}\label{eq:match:fermion:trans2}
         \vcenter{\hbox{\includegraphics[scale=1]{figs/IR_fermionfermiongluon.pdf}}} \overset{HE} \longrightarrow \vcenter{\hbox{\includegraphics[scale=1]{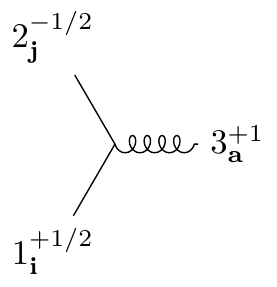}}}  & \equiv   \Omega^{\hat{i}}_{\;\;\r{i}} \; \Omega^{\hat{j}}_{\;\;\r{j}} \; \c{O}^{\hat{a}}_{\;\; \r{a}}  \left(\vcenter{\hbox{\includegraphics[scale=1]{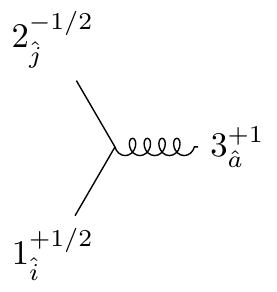}}}\right) \nn \\
      - m_\r{a} (X^{f}_{1\;\r{a}})_{\r{i}\r{j}}\c{A}(1^{+1/2},2^{-1/2},3^{+1}) & \equiv  \Omega^{\hat{i}}_{\;\;\r{i}} \; \Omega^{\hat{j}}_{\;\;\r{j}} \; \c{O}^{\hat{a}}_{\;\; \r{a}} \; (T^f_{\hat{a}})_{\hat{j}\hat{i}}    \c{A}(1^{+1/2},2^{-1/2},3^{+1})
    \end{align}

    \item $\c{A}(\b{1}^{+1/2}_\r{i},\b{2}^{+1/2}_\r{j},\b{1}^{0}_\r{a})$
    \begin{align}\label{eq:match:fermion:long}
        \vcenter{\hbox{\includegraphics[scale=1]{figs/IR_fermionfermiongluon.pdf}}} \overset{HE} \longrightarrow \vcenter{\hbox{\includegraphics[scale=1]{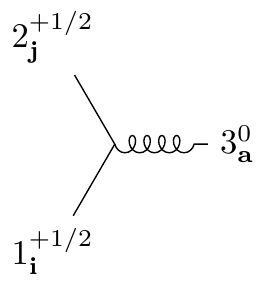}}}  & \equiv  \Omega^{\hat{i}}_{\;\;\r{i}} \; \Omega^{\hat{j}}_{\;\;\r{j}} \; \c{U}^{\hat{a}}_{\;\; I} \left(\vcenter{\hbox{\includegraphics[scale=1]{figs/UV_fermionfermionscalar.pdf}}}\right) \nn \\
      \hspace{-20pt}  \left(m_\r{j} X^{f}_{1\;\r{a}} - m_\r{i}X^{f}_{2\;\r{a}}+m_\r{a}X^{f}_{3\;\r{a}}\right)_{\r{i}\r{j}} \c{A}(1^{+1/2},2^{+1/2},3^{0}) & \equiv  \Omega^{\hat{i}}_{\;\;\r{i}} \; \Omega^{\hat{j}}_{\;\;\r{j}} \; \c{U}^{\hat{I}}_{\;\; \r{a}} \;Y_{\hat{I}\hat{i}\hat{j}}\; \c{A}(1^{+1/2},2^{+1/2},3^{0})
    \end{align}
\end{itemize}

Note that the kinematic part of the amplitudes on each side of the matching conditions are the same, as they should be. Thus, we are left with relations between the masses and color structures or charges only.

Of course, these only represent a small subset of the possible matching conditions we can obtain. Considering every possible spin configuration leads to redundant constraints, and so we omit those. However, we must address the possibility of IR amplitudes with massless particles. 

For the case of possibly massless particles in the IR, its easier to make the simultaneous replacements $\r{a}\rightarrow a,m_\r{a}\rightarrow m_{a}= 0$ and/or $\r{i}\rightarrow i,m_{\r{i}}\rightarrow m_i =  0$, than compute the actual IR amplitudes from first principles. For example, eq.~\eqref{eq:match:boson:trans} reads $ h_{{a}\r{b}\r{c}} = \fuv_{\hat{a}\hat{b}\hat{c}}\,\c{O}^{\hat{a}}_{\;\; {a}}\,\c{O}^{\hat{b}}_{\;\; \r{b}}\,\c{O}^{\hat{c}}_{\;\; \r{c}}$ for $\r{a}\rightarrow a$. Considering every possible case is not necessary and obscures the simplicity of the constraints and solutions. As such, we use we introduce an underlined notation, that can represent a massless or massive particle, summarized in Table~\ref{table:IR_notation}. With this notation, we can make the replacements to consider all combinations of massless and massive by $\r{a}\rightarrow \u{a}$, $\r{b}\rightarrow \u{b}$ and $\r{c}\rightarrow \u{c}$, where $\u{a} = \{a,\r{a}\}$, $\u{b} =\{b,\r{b}\}$ and $\u{c} = \{c,\r{c}\}$. The matching now takes the form,
\begin{itemize}
    \item $\c{A}(\b{1}^{+1}_{\u{a}},\b{2}^{+1}_{\u{b}},\b{1}^{-1}_{\u{c}})$
    \begin{align}\label{eq:match:boson:trans:all}
     \vcenter{\hbox{\includegraphics[scale=1]{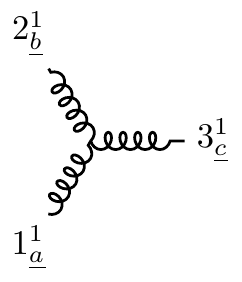}}} \overset{HE} \longrightarrow \vcenter{\hbox{\includegraphics[scale=1]{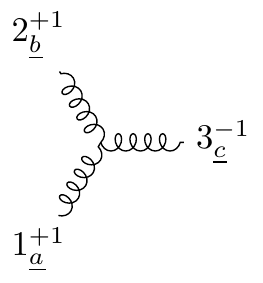}}}  & \equiv \c{O}^{\hat{a}}_{\;\; \u{a}} \c{O}^{\hat{b}}_{\;\; \u{b}} \c{O}^{\hat{c}}_{\;\; \u{c}} \left(\vcenter{\hbox{\includegraphics[scale=1]{figs/UV_gluons_abc.pdf}}}\right) \nn \\
     \fir_{\u{a}\u{b}\u{c}}\c{A}(1^{+1},2^{+1},3^{-1}) & \equiv \c{O}^{\hat{a}}_{\;\; \u{a}} \c{O}^{\hat{b}}_{\;\; \u{b}} \c{O}^{\hat{c}}_{\;\; \u{c}}\;\fuv_{\hat{a}\hat{b}\hat{c}} \c{A}(1^{+1},2^{+1},3^{-1}) 
    \end{align}

        \item $\c{A}(\b{1}^{0}_{\u{a}},\b{2}^{0}_{\u{b}},\b{1}^{-1}_{\u{c}})$
    \begin{align}\label{eq:match:boson:long:all}
         \vcenter{\hbox{\includegraphics[scale=1]{figs/IR_gluons_abc_all.pdf}}} \overset{HE} \longrightarrow \vcenter{\hbox{\includegraphics[scale=1]{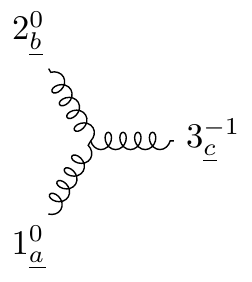}}}  & \equiv \c{U}^{\hat{I}}_{\;\; \u{a}}\;\c{U}^{\hat{J}}_{\;\; \u{b}}\;\c{O}^{\hat{a}}_{\;\; \u{c}} \left(\vcenter{\hbox{\includegraphics[scale=1]{figs/UV_scalarscalargluon.pdf}}}\right) \nn \\
         \fir_{\u{a}\u{b}\u{c}}\frac{(m_{\u{c}}^2 - m_{\u{a}}^2 -m_{\u{b}}^2)}{2\,m_{\u{a}}m_{\u{b}}}\c{A}(1^{0},2^{0},3^{-1}) & \equiv \c{U}^{\hat{I}}_{\;\; \u{a}}\;\c{U}^{\hat{J}}_{\;\; \u{b}}\;\c{O}^{\hat{a}}_{\;\; \u{c}}\;(T^s_{\hat{a}})_{\hat{I}\hat{J}} \c{A}(1^{0},2^{0},3^{-1})
    \end{align}

    \item $\c{A}(\b{1}^{+1/2}_{\u{i}},\b{2}^{+1/2}_{\u{j}},\b{1}^{+1}_{\u{a}})$
    \begin{align}\label{eq:match:fermion:allplus:all}
         \vcenter{\hbox{\includegraphics[scale=1]{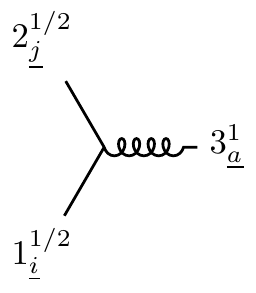}}} \overset{HE} \longrightarrow \vcenter{\hbox{\includegraphics[scale=1]{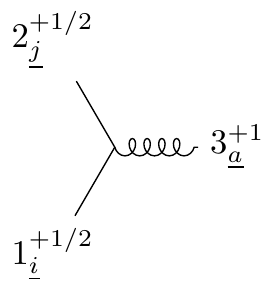}}}  & \equiv  0 \nn \\
        (X^{f}_{3\;\u{a}})_{\u{i}\u{j}}\c{A}(1^{+1/2},2^{+1/2},3^{+1}) & \equiv  0_{\u{a}\u{i}\u{j}}
    \end{align}

    \item $\c{A}(\b{1}^{-1/2}_{\u{i}},\b{2}^{-1/2}_{\u{j}},\b{1}^{-1}_{\u{a}})$
    \begin{align}\label{eq:match:fermion:allminus:all}
         \vcenter{\hbox{\includegraphics[scale=1]{figs/IR_fermionfermiongluon_all.pdf}}} \overset{HE} \longrightarrow \vcenter{\hbox{\includegraphics[scale=1]{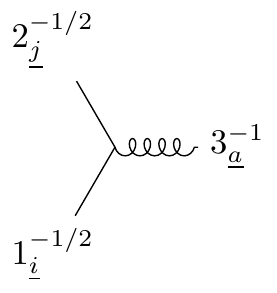}}}  & \equiv  0 \nn \\
        (X^{f}_{4\;\u{a}})_{\u{i}\u{j}}\c{A}(1^{-1/2},2^{-1/2},3^{-1}) & \equiv  0_{\u{a}\u{i}\u{j}}
    \end{align}

    \item $\c{A}(\b{1}^{-1/2}_{\u{i}},\b{2}^{+1/2}_{\u{j}},\b{1}^{+1}_{\u{a}})$
    \begin{align}\label{eq:match:fermion:trans1:all}
         \vcenter{\hbox{\includegraphics[scale=1]{figs/IR_fermionfermiongluon_all.pdf}}} \overset{HE} \longrightarrow \vcenter{\hbox{\includegraphics[scale=1]{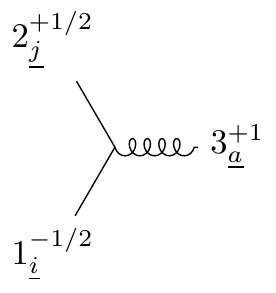}}}  & \equiv  \Omega^{\hat{i}}_{\;\;\u{i}} \; \Omega^{\hat{j}}_{\;\;\u{j}} \; \c{O}^{\hat{a}}_{\;\; \u{a}}  \left(\vcenter{\hbox{\includegraphics[scale=1]{figs/UV_fermionfermiongluon_trans1.pdf}}}\right) \nn \\
       m_{\u{a}} (X^{f}_{2\;\u{a}} +X^{f}_{3\;\u{a}})_{\u{i}\u{j}}\c{A}(1^{-1/2},2^{+1/2},3^{+1}) & \equiv  \Omega^{\hat{i}}_{\;\;\u{i}} \; \Omega^{\hat{j}}_{\;\;\u{j}} \; \c{O}^{\hat{a}}_{\;\; a} (T^f_{\hat{a}})_{\hat{i}\hat{j}}  \c{A}(1^{-1/2},2^{+1/2},3^{+1}) 
    \end{align}

    \item $\c{A}(\b{1}^{+1/2}_{\u{i}},\b{2}^{-1/2}_{\u{j}},\b{1}^{+1}_{\u{a}})$
    \begin{align}\label{eq:match:fermion:trans2:all}
         \vcenter{\hbox{\includegraphics[scale=1]{figs/IR_fermionfermiongluon_all.pdf}}} \overset{HE} \longrightarrow \vcenter{\hbox{\includegraphics[scale=1]{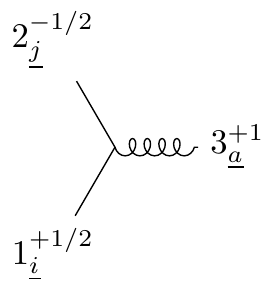}}}  & \equiv   \Omega^{\hat{i}}_{\;\;\u{i}} \; \Omega^{\hat{j}}_{\;\;\r{j}} \; \c{O}^{\hat{a}}_{\;\; \u{a}}  \left(\vcenter{\hbox{\includegraphics[scale=1]{figs/UV_fermionfermiongluon_trans2.pdf}}}\right) \nn \\
      - m_{\u{a}} (X^{f}_{1\;\u{a}})_{\u{i}\u{j}}\c{A}(1^{+1/2},2^{-1/2},3^{+1}) & \equiv  \Omega^{\hat{i}}_{\;\;\u{i}} \; \Omega^{\hat{j}}_{\;\;\u{j}} \; \c{O}^{\hat{a}}_{\;\; \u{a}} \; (T^f_{\hat{a}})_{\hat{j}\hat{i}}    \c{A}(1^{+1/2},2^{-1/2},3^{+1})
    \end{align}

    \item $\c{A}(\b{1}^{+1/2}_{\u{i}},\b{2}^{+1/2}_{\u{j}},\b{1}^{0}_{\u{a}})$
    \begin{align}\label{eq:match:fermion:long:all}
        \vcenter{\hbox{\includegraphics[scale=1]{figs/IR_fermionfermiongluon_all.pdf}}} \overset{HE} \longrightarrow \vcenter{\hbox{\includegraphics[scale=1]{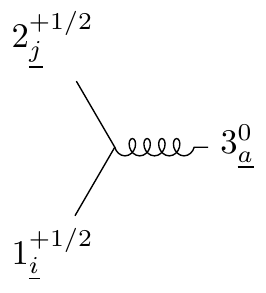}}}  & \equiv  \Omega^{\hat{i}}_{\;\;\u{i}} \; \Omega^{\hat{j}}_{\;\;\u{j}} \; \c{U}^{\hat{a}}_{\;\; \hat{I}} \left(\vcenter{\hbox{\includegraphics[scale=1]{figs/UV_fermionfermionscalar.pdf}}}\right) \nn \\
      \hspace{-20pt}  \left(m_{\u{j}} X^{f}_{1\;\u{a}} - m_{\u{i}}X^{f}_{2\;\u{a}}+m_{\u{a}}X^{f}_{3\;\u{a}}\right)_{\u{i}\u{j}} \c{A}(1^{+1/2},2^{+1/2},3^{0}) & \equiv  \Omega^{\hat{i}}_{\;\;\u{i}} \; \Omega^{\hat{j}}_{\;\;\u{j}} \; \c{U}^{\hat{I}}_{\;\; \u{a}} \;Y_{\hat{I}\hat{i}\hat{j}}\; \c{A}(1^{+1/2},2^{+1/2},3^{0})
    \end{align}
\end{itemize}

We summarize ALL the constraints from matching below,
\begin{align}
  h_{\u{a}\u{b}\u{c}} &= \fuv_{\hat{a}\hat{b}\hat{c}}\,\c{O}^{\hat{a}}_{\;\; \u{a}}\,\c{O}^{\hat{b}}_{\;\; \u{b}}\,\c{O}^{\hat{c}}_{\;\; \u{c}} \label{eq:constraint:1}\,,\\
  \fir_{\u{a}\u{b}\u{c}}\frac{(m_{\u{c}}^2 - m_{\u{a}}^2 -m_{\u{b}}^2)}{2\,m_{\u{a}}m_{\u{b}}}&= \c{U}^{\hat{I}}_{\;\; \u{a}}\;\c{U}^{\hat{J}}_{\;\; \u{b}}\;\c{O}^{\hat{c}}_{\;\; \u{c}}\;(T^s_{\hat{c}})_{\hat{I}\hat{J}}\label{eq:constraint:2} \,,\\
   m_{\u{a}} (X^{f}_{2\;\u{a}})_{\u{i}\u{j}} & =  \Omega^{\hat{i}}_{\;\;\u{i}} \; \Omega^{\hat{j}}_{\;\;\u{j}} \; \c{O}^{\hat{a}}_{\;\; \u{a}} (T^f_{\hat{a}})_{\hat{i}\hat{j}} \label{eq:constraint:3}\,, \\
  - m_{\u{a}} (X^{f}_{1\;\u{a}})_{\u{i}\u{j}} & =  \Omega^{\hat{i}}_{\;\;\u{i}} \; \Omega^{\hat{j}}_{\;\;\u{j}} \; \c{O}^{\hat{a}}_{\;\; \u{a}} \; (T^f_{\hat{a}})_{\hat{j}\hat{i}} \label{eq:constraint:4}  \,,\\
   m_{\u{j}} (X^{f}_{1\;\u{a}})_{\u{i}\u{j}} - m_{\u{i}} (X^{f}_{2\;\u{a}})_{\u{i}\u{j}} & = \Omega^{\hat{i}}_{\;\;\u{i}} \; \Omega^{\hat{j}}_{\;\;\u{j}} \; \c{U}^{\hat{I}}_{\;\; \u{a}} \;Y_{\hat{I}\hat{i}\hat{j}}\label{eq:constraint:5}\,.
\end{align}

\subsection{Matching Solutions\label{sec:MatchingSolutions}}

We will first present the solutions to the above constraints, and then show why they work.
The solutions are,
\begin{itemize}
    \item The structure constants in the IR are simply a change of basis or rotation of the structure constants in the UV,
    \begin{align}\label{eq:solution1}
      h_{\u{a}\u{b}\u{c}} &= \fuv_{\hat{a}\hat{b}\hat{c}}  \c{O}^{\hat{a}}_{\;\; \u{a}} \c{O}^{\hat{b}}_{\;\; \u{b}} \c{O}^{\hat{c}}_{\;\; \u{c}} \,,
    \end{align}
    for $\u{a} = \{a,\r{a}\}$, $\u{b} =\{b,\r{b}\}$ and $\u{c} = \{c,\r{c}\}$. In other words, the couplings of any three massive or massless spin-$1$ particles in the IR can be determined from their massless counterparts in the UV.

    \item The `generators' in the IR, in the scalar and adjoint representation, are simply a change of basis of the generators in the UV,
    \begin{align}\label{eq:solution2}
      X^s_{\u{a}} = T^s_{\hat{a}} \c{O}^{\hat{a}}_{\;\; \u{a}} \quad \text{and} \quad X^{\text{ad}}_{\u{a}} &= T^{\text{ad}}_{\hat{a}} \c{O}^{\hat{a}}_{\;\; \u{a}} \,,
    \end{align}
    for $\u{a} = \{a,\r{a}\}$. So the symmetry group, or more specifically the algebra, in the UV is inherited by the IR.

    \item The masses of the bosons are generated by
    \begin{align}\label{eq:solution3}
      m_{\u{a}} U^{\hat{I}}_{\;\;\u{a}} = (X^s_{\u{a}})^{\hat{I}\hat{J}}v_{\hat{J}} = (T^s_{\hat{a}} \c{O}^{\hat{a}}_{\;\; \u{a}})^{\hat{I}\hat{J}}v_{\hat{J}} \,,
    \end{align}
    for $\u{a} = \{a,\r{a}\}$ and for some vector $\vec{v}$. Note that $\vec{v}$ here plays a similar role to that of the vacuum expectation value in the traditional Higgs mechanism, however, in our case, we have no mention of quantum fields. Further, since $U$ is an element of $SO(N_s)$, then
    \begin{align}
        m_a^2\delta_{ab} = \vec{v}^TX^{sT}_aX^s_b\vec{v}\,.
    \end{align}
    
    \item The masses of the fermions are generated by
    \begin{align}\label{eq:solution4}
        \text{diag}(m_1,\dots,m_{N_f}) =  m_{\u{k}} \delta^{\;\u{i}}_{\u{k}} &=  [\Omega^T (v^{\hat{I}}Y_{\hat{I}\hat{j}\hat{k}} + \tilde{m}_{\hat{j}\hat{k}})\Omega]_{\u{k}}^{\;\u{i}} \,,
    \end{align}
    for $m_i$ real and non-negative, and $\tilde{m}_{jk}$ contributions from a background scalar, eq.~\eqref{eq:yukawa:constraint:gauge:mass}. More succinctly, $\text{diag}(m_1,\dots,m_{N_f}) = \Omega^T (\vec{v}.{Y} + \tilde{m})\Omega$. See \cite{Dreiner:2008tw} for discussions on Takagi diagonalization. Note that non-zero gauge invariant fermion masses are allowed, with no contributions from the Yukawa couplings or background scalars, which we highlight in Section~\ref{sec:discussion}.
\end{itemize}

For the rest of this subsection, we will show why these are the solutions. The approach will be to substitute our solutions into our constraints and use our conditions from consistent factorization.

The first solution, eq.~\eqref{eq:solution1} is the easiest, as it is a consequence of considering all the combinations of massive and massless particles in the IR from eq.~\eqref{eq:constraint:1}.

The second and third solution, eq.~\eqref{eq:solution2} and eq.~\eqref{eq:solution3}, solve the constraint eq.~\eqref{eq:constraint:2}.
To illustrate, starting with eq.~\eqref{eq:constraint:2},
\begin{align}\label{eq:app:constraint_gauge_boson}
  \frac{1}{2}\,\fir_{\u{a}\u{b}\u{c}}(m_{\u{c}}^2 - m_{\u{a}}^2 -m_{\u{b}}^2)&= m_{\u{a}}\,\c{U}^{\hat{I}}_{\;\; \u{a}}\,m_{\u{b}}\,\c{U}^{\hat{J}}_{\;\; \u{b}}\;\c{O}^{\hat{c}}_{\;\; \u{c}}\;(T^s_{\hat{c}})_{\hat{I}\hat{J}}\,,
\end{align}
we will consider the right hand side of the equation, substitute the solution, and show that it is equal to the left hand side. Let us proceed.
Substituting the solution on the right hand side, and using the anti-symmetry of $(T^s_{\hat{a}})_{\hat{I}\hat{J}} = - (T^s_{\hat{a}})_{\hat{J}\hat{I}}$,
\begin{align} \label{eq:boson:proof1}
  m_{\u{a}}\,\c{U}^{\hat{I}}_{\;\; \u{a}}\,m_{\u{b}}\,\c{U}^{\hat{J}}_{\;\; \u{b}}\;\c{O}^{\hat{c}}_{\;\;\u{c}}\;(T^s_{\hat{c}})_{\hat{I}\hat{J}} &= (X^s_{\u{a}})^{\hat{I}\hat{K}}v_{\hat{K}} \,(T^s_{\hat{c}}\c{O}^{\hat{c}}_{\;\; \u{c}})_{\hat{I}\hat{J}} \, (X^s_{\u{b}})^{\hat{J}\hat{L}}v_{\hat{L}}\nonumber \\
  &=\c{O}^{\hat{a}}_{\;\; {\u{a}}}\,\c{O}^{\hat{b}}_{\;\; {\u{b}}}\,\c{O}^{\hat{c}}_{\;\; {\u{b}}}\,v_{\hat{K}}\,[(T^s_{\hat{a}})^{\hat{I}\hat{K}}\,(T^s_{\hat{c}})_{\hat{I}\hat{J}}\,(T^s_{\hat{b}})^{\hat{J}\hat{L}}]\,v_{\hat{L}} \nonumber \\
  &=\c{O}^{\hat{a}}_{\;\; {\u{a}}}\,\c{O}^{\hat{b}}_{\;\; {\u{b}}}\,\c{O}^{\hat{c}}_{\;\; {\u{c}}}\,v_{\hat{K}}\,[-(T^s_{\hat{a}})^{\hat{K}\hat{I}}\,(T^s_{\hat{c}})_{\hat{I}\hat{J}}\,(T^s_{\hat{b}})^{\hat{J}\hat{L}}]\,v_{\hat{L}} \nonumber \\
  &=\c{O}^{\hat{a}}_{\;\; {\u{a}}}\,\c{O}^{\hat{b}}_{\;\; {\u{b}}}\,\c{O}^{\hat{c}}_{\;\; {\u{c}}}\,v_{\hat{L}}\,[(T^s_{\hat{a}})^{\hat{L}\hat{J}}\,(T^s_{\hat{c}})_{\hat{J}\hat{I}}\,(T^s_{\hat{b}})^{\hat{I}\hat{K}}]\,v_{\hat{K}} \nonumber \\
  &=\frac{1}{2}\c{O}^{\hat{a}}_{\;\; {\u{a}}}\,\c{O}^{\hat{b}}_{\;\; {\u{b}}}\,\c{O}^{\hat{c}}_{\;\; {\u{c}}}\,v_{\hat{K}}\,[T^s_{\hat{b}}T^s_{\hat{c}}T^s_{\hat{a}}-T^s_{\hat{a}}T^s_{\hat{b}}T^s_{\hat{c}}]^{\hat{K}\hat{L}}\,v_{\hat{L}} \nonumber \\
  &=\frac{1}{2}v_{\hat{K}}\,[X^s_{\u{b}}X^s_{\u{c}}X^s_{\u{a}}-X^s_{\u{a}}X^s_{\u{b}}X^s_{\u{c}}]^{\hat{K}\hat{L}}\,v_{\hat{L}} \,.
\end{align}
Next, by using the commutation relation from consistent factorization in eq.~\eqref{eq:yangmillstructure}, $[T_{\hat{a}},T_{\hat{b}}] = f_{\hat{a}\hat{b}}^{\;\;\;\;\hat{c}}\;T_{\hat{c}}$ repeatedly, we have,
\begin{align}
  T_{\hat{b}}T_{\hat{c}}T_{\hat{a}} &= (T_{\hat{c}}T_{\hat{b}} + f_{\hat{b}\hat{c}}^{\;\;\;\;\hat{d}}\, T_{\hat{d}})\, T_{\hat{a}} \nonumber \\
  &= T_{\hat{c}}(T_{\hat{a}}T_{\hat{b}} + f_{\hat{b}\hat{a}}^{\;\;\;\;\hat{d}}\, T_{\hat{d}})+ f_{\hat{b}\hat{c}}^{\;\;\;\;\hat{d}}\, T_{\hat{d}} T_{\hat{a}} \nonumber \\
  &= (T_{\hat{a}}T_{\hat{c}} + f_{\hat{c}\hat{a}}^{\;\;\;\;\hat{d}}\,T_{\hat{d}})\, T_{\hat{b}} +f_{\hat{b}\hat{a}}^{\;\;\;\;\hat{d}}\, T_{\hat{c}} T_{\hat{d}} + f_{\hat{b}\hat{c}}^{\;\;\;\;\hat{d}}\, T_{\hat{d}} T_{\hat{a}} \nonumber \\
  &= T_{\hat{b}}T_{\hat{c}}T_{\hat{a}} +f_{\hat{c}\hat{a}}^{\;\;\;\;\hat{d}}\, T_{\hat{d}} T_{\hat{b}} + f_{\hat{b}\hat{a}}^{\;\;\;\;\hat{d}}\, T_{\hat{c}} T_{\hat{d}} + f_{\hat{b}\hat{c}}^{\;\;\;\;\hat{d}}\, T_{\hat{d}} T_{\hat{a}} \nonumber \\
  T_{\hat{b}}T_{\hat{c}}T_{\hat{a}} -T_{\hat{b}}T_{\hat{c}}T_{\hat{a}} &=  f_{\hat{c}\hat{a}}^{\;\;\;\;\hat{d}}\, T_{\hat{d}} T_{\hat{b}} + f_{\hat{b}\hat{a}}^{\;\;\;\;\hat{d}}\, T_{\hat{c}} T_{\hat{d}} + f_{\hat{b}\hat{c}}^{\;\;\;\;\hat{d}}\, T_{\hat{d}} T_{\hat{a}} \,.
\end{align}
Acting on the above with the rotations $\c{O}^{\hat{a}}_{\;\; {\u{a}}}\,\c{O}^{\hat{b}}_{\;\; {\u{b}}}\,\c{O}^{\hat{c}}_{\;\; {\u{c}}}$
\begin{align} 
  X_{\u{b}}\, X_{\u{c}} \, X_{\u{a}} - X_{\u{a}} \, X_{\u{c}} \, X_{\u{b}} &= h_{{\u{c}}{\u{a}}}^{\;\;\;\;{\u{d}}}\,X_{\u{d}} \, X_{\u{b}} + h_{{\u{b}}{\u{a}}}^{\;\;\;\;{\u{d}}}\,X_{\u{c}}\,X_{\u{d}} + h_{{\u{b}}{\u{c}}}^{\;\;\;\;{\u{d}}}\,X_{\u{d}}\,X_{\u{a}} \,,
\end{align}
and contracting with $v$, eq.~\eqref{eq:boson:proof1} now reads,
\begin{align}\label{eq:boson:proof2}
  \frac{1}{2}\vec{v}[X_{\u{b}}^s\, X_{\u{c}}^s \, X_{\u{a}}^s - X_{\u{a}}^s \, X_{\u{c}}^s \, X_{\u{b}}^s]\vec{v} &= \frac{1}{2}v_{\hat{K}}[h_{{\u{c}}{\u{a}}}^{\;\;\;\;{\u{d}}}\,X^s_{\u{d}} \, X^s_{\u{b}} + h_{{\u{b}}{\u{a}}}^{\;\;\;\;{\u{d}}}\,X^s_{\u{c}}\,X^s_{\u{d}} + h_{{\u{b}}{\u{c}}}^{\;\;\;\;{\u{d}}}\,X^s_{\u{d}}\,X^s_{\u{a}} ]^{\hat{K}\hat{L}} v_{\hat{L}} \,.
\end{align}

Now, consider the term $h_{{\u{c}}{\u{a}}}^{\;\;\;\;{\u{d}}}\,v_{\hat{K}}[X_{\u{d}} \, X_{\u{b}}]^{\hat{K}\hat{L}}v_{\hat{L}}$ in the eq above,
\begin{align}
  h_{{\u{c}}{\u{a}}}^{\;\;\;\;{\u{d}}}\,v_{\hat{K}}[X_{\u{d}} \, X_{\u{d}}]^{\hat{K}\hat{L}}v_{\hat{L}} &= h_{{\u{a}}{\u{c}}}^{\;\;\;\;{\u{d{\u{c}}}}}\,V_{\hat{K}} (X_{\u{d}})^{\hat{K}}_{\;\;\hat{J}}\,(X_{\u{b}})^{\hat{J}\hat{L}}\,V_{\hat{L}}  \nonumber \\
  &= - h_{{\u{c}}{\u{a}}}^{\;\;\;\;{\u{d}}}\,(X_{\u{d}})_{\hat{J}}^{\;\;\hat{K}} v_{\hat{K}}\,(X_{\u{b}})^{\hat{J}\hat{L}}\,v_{\hat{L}} \,.
\end{align}
This can be simplified again by inserting our solution,
\begin{align}
  h_{{\u{c}}{\u{a}}}^{\;\;\;\;{\u{d}}}\,v_{\hat{K}}[X_{\u{d}} \, X_{\u{b}}]^{\hat{K}\hat{L}}v_{\hat{L}} &= - h_{{\u{c}}{\u{a}}}^{\;\;\;\;{\u{d}}} \,m_{\u{d}}\,U_{{\u{d}}\hat{J}}\,U_{\u{b}}^{\;\;\hat{J}}\,m_{\u{b}} \nonumber \\
  &=- h_{{\u{c}}{\u{a}}}^{\;\;\;\;{\u{d}}} \,m_{\u{d}}\,(U^TU)_{{\u{d}}{\u{b}}}\,m_{\u{b}} \nonumber \\
  &=-h_{{\u{c}}{\u{a}}{\u{b}}}\,m_{\u{b}}^2 \,.
\end{align}
So, making the replacements above in eq.~\eqref{eq:boson:proof2},
\begin{align}
  \frac{1}{2}\vec{v}[X_{\u{b}}^s\, X_{\u{c}}^s \, X_{\u{a}}^s - X_{\u{a}}^s \, X_{\u{c}}^s \, X_{\u{b}}^s]\vec{v} &= \frac{1}{2}(-h_{{\u{c}}{\u{a}}{\u{b}}}\,m_{\u{b}}^2 -h_{{\u{b}}{\u{a}}{\u{c}}}\,m_{\u{c}}^2 -h_{{\u{b}}{\u{a}}{\u{c}}}\,m_{\u{a}}^2) \nonumber \\
  &= \frac{1}{2}h_{{\u{a}}{\u{b}}{\u{c}}}\,(m_{\u{c}}^2 - m_{\u{b}}^2 - m_{\u{a}}^2)\,,
\end{align}
which is the left hand side of eq.~\eqref{eq:app:constraint_gauge_boson}, which we set out to show.

Finally, we show that fermion mass in eq.~\eqref{eq:solution4} is a solution. To do this, we will take our constraints, substitute our solutions, and then show that the resulting equation is that one would expect from consistent factorization from eq.~\eqref{eq:yukawa:constraint}.

We begin by combining our constraints eq.~\eqref{eq:constraint:3} and eq.~\eqref{eq:constraint:4} in eq.~\eqref{eq:constraint:5}, and making the replacements to consider all combinations of massive and massless by $\r{a}\rightarrow \u{a}$, $\r{b}\rightarrow \u{b}$ and $\r{c}\rightarrow \u{c}$, where $\u{a} = \{a,\r{a}\}$, $\u{b} =\{b,\r{b}\}$ and $\u{c} = \{c,\r{c}\}$, we have,
\begin{align}
      (X^{f}_{1\;\u{a}})_{\u{i}\u{j}}\,m_{\u{j}}  - (X^{f}_{2\;\u{a}})_{\u{i}\u{j}}\, m_{\u{i}}& =  \Omega^{\hat{i}}_{\;\;\u{i}} \; \Omega^{\hat{j}}_{\;\;\u{j}} \; \c{U}^{\hat{I}}_{\;\; \u{a}} \;Y^f_{\hat{I}\hat{i}\hat{j}}\,,
\end{align}
which, after substitution of eq.~\eqref{eq:constraint:3} and eq.~\eqref{eq:constraint:4} give,
\begin{align}
  m_{\u{j}}\,(\Omega^T)_{\u{i}}^{\;\;\hat{i}}\,(T^f_{\hat{a}}\,\c{O}^{\hat{a}}_{\;\; {\u{a}}})_{\hat{j}\hat{i}}\,\Omega_{\;\;{\u{j}}}^{\hat{j}}  +  m_{\u{i}}\,(\Omega^T)_{\u{j}}^{\;\;\hat{j}}\,(T^f_{\hat{a}} \c{O}^{\hat{a}}_{\;\; {\u{a}}})_{\hat{i}\hat{j}}\,\Omega_{\;\;{\u{i}}}^{\hat{i}} + (\Omega^T)_{\u{j}}^{\;\;\hat{j}}\,Y^f_{\hat{I}\hat{j}\hat{k}}\,U^{\hat{I}}_{\;\;{\u{a}}}\,m_{\u{a}}\,\Omega_{\;\;{\u{i}}}^{\hat{i}} &= 0 \,.
\end{align}

Substituting for the Higgs mechanism eq.~\eqref{eq:solution3}, and removing some index contractions, we have,
\begin{align}\label{eq:fermion:proof:a}
  m_{\u{j}}\,\delta_{\u{j}}^{\;{\u{k}}}[\Omega^T (T^f\c{O})_{\u{a}}\Omega]_{{\u{k}}{\u{i}}} + m_{\u{i}}\,\delta_{\u{i}}^{\;{\u{k}}}[\Omega^T (T^f\c{O})_{{\u{a}}}\Omega]_{{\u{k}}{\u{j}}} + [\Omega^T (Y^f.X^s.\vec{v})_{{\u{a}}}\Omega]_{{\u{i}}{\u{j}}} &= 0 \,.
  \end{align}
(no sum over $i$ and $j$). Substituting the solution for fermion masses eq.~\eqref{eq:solution4}, and the relation between the UV and IR generators in the representation of the scalars,
\begin{align}
    [ \Omega^T(\vec{v}.Y+\tilde{m})\Omega \Omega^T (T^f\c{O})_{\u{a}}\Omega\ ]_{{\u{i}}{\u{j}}} + [ \Omega^T(\vec{v}.Y+\tilde{m})\Omega \Omega^T (T^f\c{O})_{{\u{a}}}\Omega]_{{\u{j}}{\u{i}}} + [\Omega^T (Y.T^s\c{O}.\vec{v})_{{\u{a}}}\Omega]_{{\u{i}}{\u{j}}} &= 0 \,.
\end{align} 
Multiplying from the left by $\Omega$ and the right by $\Omega^T$ and $\c{O}^{-1}$, then making the obvious cancellations $\Omega \Omega^T = \mathrm{1}$ yields,
\begin{align}
    (\vec{v}.Y+\tilde{m})T^f +  (\vec{v}.Y+\tilde{m})T^f + (Y.T^s.\vec{v}) &= 0 \,,
\end{align}
which is simply the sum of the Yukawa constraint eq.~\eqref{eq:yukawa:constraint}
\begin{align}
       (\vec{v}.Y)T^f +  (\vec{v}.Y)T^f + (Y.T^s.\vec{v}) &= 0  \,, 
\end{align}
and the background scalar mass eq.~\eqref{eq:yukawa:constraint:gauge:mass}
\begin{align}
    \tilde{m}T^f +   \tilde{m}T^f  = 0 \,.
\end{align}
Hence, we have shown that these are the solutions to our constraints.

\subsection{Discussion \label{sec:discussion}}

Consider eq.~\eqref{eq:solution3} for massless particles in the IR $\u{a} = a$ and $m_{\u{a}} = 0$,
\begin{align}\label{eq:massless:spin:1}
  0_{\hat{J}} &= v^{\hat{I}}(X^s_{{a}})_{\hat{I}\hat{J}} = v^{\hat{I}}(T^s_{\hat{a}} \c{O}^{\hat{a}}_{\;\; {a}})_{\hat{I}\hat{J}} \,.
\end{align}
This is the statement that for every massless spin-$1$ particle in the IR, $a$, there is a generator $X^s_a$ that annihilates $\vec{v}$. This places constraints on the form of $\vec{v}$. 

Next, consider two massless spin-$1$ particles in the IR, $a$ and $b$. By the above, $(X_a^s)\vec{v} = \vec{0}$ and $(X_b^s) \vec{v} = \vec{0}$. So, $[X_a^s,X_b^s]\vec{v} = \vec{0}$. But our solutions eq.~\eqref{eq:solution1} and \eqref{eq:solution2} also tell us that $[X_a^s,X_b^s] = h_{ab}^{\;\;\;\;\u{c}} X^s_{\u{c}}$. Acting on $\vec{v}$, 
\begin{align}
    [X_a^s,X_b^s]\vec{v} &= h_{ab}^{\;\;\;\;\u{c}} X^s_{\u{c}}\vec{v} \nn \\
   \vec{0} &= h_{ab}^{\;\;\;\;{c}} X^s_{{c}}\vec{v} + h_{ab}^{\;\;\;\;\r{c}} X^s_{\r{c}}\vec{v}\nn \\
   \vec{0} &= \vec{0} +  h_{ab}^{\;\;\;\;\r{c}} X^s_{\r{c}}\vec{v}\,.
\end{align}
However, for massive particles in the IR, $\r{c}$, $X^s_{\r{c}} \vec{v} \neq \vec{0}$ in general, and so we conclude that $ h_{ab}^{\;\;\;\;\r{c}} X^s_{\r{c}} = 0_{ab}$. This means that coupling constant between two massless spin-$1$ particles and one massive spin-$1$ particle is zero, or, a massive spin-$1$ particle cannot decay to two massless spin-$1$ particles. Of course, we already independently knew this from Yang's theorem, which is a statement from  a kinematics. It was amusing to see it here from pure group theory.

With this in mind, we can revisit the commutator,
\begin{align}
    [X_a^s,X_b^s] &= h_{ab}^{\;\;\;\;\u{c}} X^s_{\u{c}} \nn\\
   &= h_{ab}^{\;\;\;\;{c}} X^s_{{c}} + h_{ab}^{\;\;\;\;\r{c}} X^s_{\r{c}}\nn\\ 
   &=    h_{ab}^{\;\;\;\;{c}} X^s_{{c}} \,,
\end{align}
which says that the generators associated with massless particles form a subalgebra $\mathfrak{h}$. So we have discovered that, if there are massless particles in the IR, then there must be a subalgebra $\mathfrak{h}$ of $\mathfrak{g}$. Hence, there must be a subgroup $H \subset G$.

Lastly, we mention that the constraints eq.~\eqref{eq:app:constraint_gauge_boson} and eq.~\eqref{eq:fermion:proof:a} provide additional information connecting the charges (or more generally the generators) to the masses. For the simplest examples, consider an unbroken generator in the IR, which corresponds to a massless spin-$1$ particle $a$. Then, with $m_{a} = 0$, eq.~\eqref{eq:app:constraint_gauge_boson} reads,
\begin{align}
    \frac{1}{2}\,\fir_{a\u{b}\u{c}}(m_{\u{c}}^2  -m_{\u{b}}^2) &= 0\,.
\end{align}
For $\u{b} = b$ massless, $m_{\u{b}} = m_{{b}} = 0$, and the equation is almost spoiled, except for the fact that $h_{ab\u{c}} = 0$. For  $\u{b} = \r{b}$ and $\u{c} = \r{c}$ massive, then the equation reads,
\begin{align}
    \frac{1}{2}\,\fir_{a\r{b}\r{c}}(m_{\r{c}}^2  -m_{\r{b}}^2) &= 0\,.
\end{align}
and so we have $m_{\r{b}} = m_{\r{c}}$. In the SM, we know this as the masses of the $W^\pm$ being the same. 

With respect to the fermion masses, for a unbroken generator $\u{a} = a$ in the IR, eq.~\eqref{eq:fermion:proof:a} reads,
\begin{align}
  m_{\u{j}}[\Omega^T (T^f\c{O})_{{a}}\Omega]_{{\u{j}}{\u{i}}} + m_{\u{i}}[\Omega^T (T^f\c{O})_{{{a}}}\Omega]_{{\u{i}}{\u{j}}}  &= 0\,,
\end{align}
which shows that there is a possibility for fermions to have gauge invariant masses. Furthermore, since these masses do not come from Yukawa couplings or background scalars, they can exist in the UV.






\section{Standard Model \label{sec:StandardModel}}

\subsection{The UV}

The symmetry group in the UV, is $G = SU(2)_L \times U(1)_Y$ (we will be working with singlets of $SU(3)$). The Lie algebra associated with this group is spanned by four generators, three from $SU(2)_L$ and one from $U(1)_Y$,
\begin{align}
     \tilde{\mathfrak{g}} &= \left\{ \tilde T_{1_1}, \tilde T_{2_1} ,\tilde T_{3_1} ,\tilde T_{1_2} \right\} \,.
\end{align} 
The particle labels are usually denoted by $1_1 = W_1$, $1_2 = W_2$, $1_3 = W_3$ and $2_1 = B$.
As previously mentioned, we find it convenient to rescale our Lie algebra by the coupling constants,
\begin{align}
   \mathfrak{g} &= \left\{ g_1 \tilde T_{1_1}, g_1 \tilde T_{2_1}, g_1 \tilde T_{3_1}, g_2 \tilde T_{2_1}  \right\} \,,\nn \\
   &= \{  T_1,  T_2, T_3, T_4 \}\,,\nonumber \\
   &= \{  T_{W_1},  T_{W_2}, T_{W_3}, T_{B} \}\,.
\end{align}
Now that we have fixed the group structure, we can move on to defining the representations which the spin-$0$, spin-$1/2$ and spin-$1$ transform under.

\begin{figure}[h]
\centering
\includegraphics[width=0.4\textwidth]{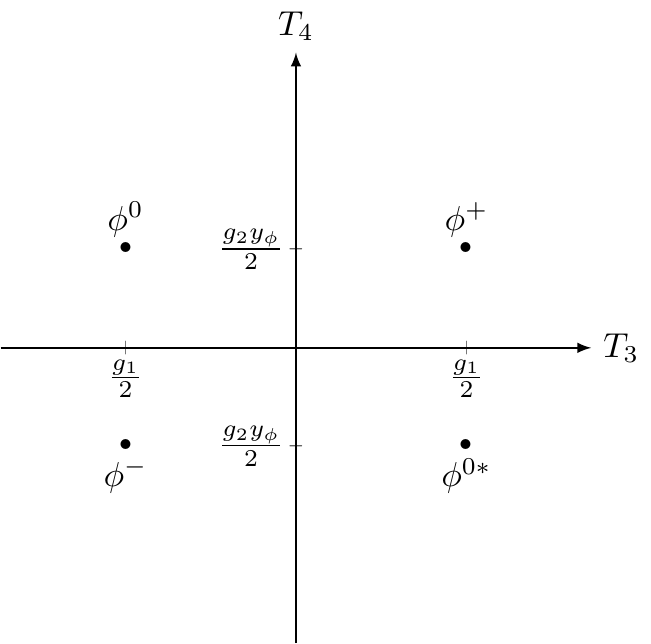} \quad
\includegraphics[width=0.4\textwidth]{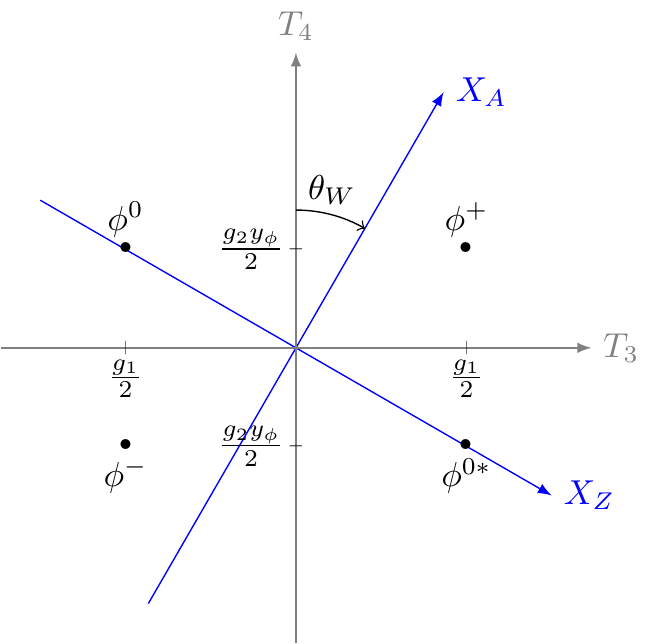}
\caption{The weight diagram of the scalar sector of the Standard Model.\label{fig:weights:higgs}}
\end{figure}

Starting with the scalar sector, we need to define $T^s_a$. The Higgs is defined as a hypercharge $Y=y_\phi$ (usually $y_\phi=1$ with our convention, though we will leave it unspecified for now), $SU(2)$ complex doublet.
The Higgs has four real degrees of freedom, and, although we can work in a basis that makes these manifest, for now we choose to work in a basis that has the maximal set of diagonal generators, which is a reducible complex representation. From Appendix~\ref{appendix:repofHiggs}, we have $(T^s_a)_{\hat{I}\hat{J}}$,
\begin{alignat}{4}\label{eq:higgs:scalar:rep:example}
T^s_1 &= \frac{g_1}{2}\left(
\begin{array}{cccc}
 0 & 1 & 0 & 0 \\
 1 & 0 & 0 & 0 \\
 0 & 0 & 0 & -1 \\
 0 & 0 & -1 & 0 \\
\end{array}
\right)\,,&\quad T^s_2 &= \frac{g_1}{2} \left(
\begin{array}{cccc}
 0 & -i & 0 & 0 \\
 i & 0 & 0 & 0 \\
 0 & 0 & 0 & -i \\
 0 & 0 & i & 0 \\
\end{array}
\right), \nonumber
\\
T^s_3 &= \frac{g_1}{2}\left(
\begin{array}{cccc}
 1 & 0 & 0 & 0 \\
 0 & -1 & 0 & 0 \\
 0 & 0 & -1 & 0 \\
 0 & 0 & 0 & 1 \\
\end{array}
\right),&\quad
T^s_4 &= \frac{g_2y_\phi}{2}  \left(
\begin{array}{cccc}
 1 & 0 & 0 & 0 \\
 0 & 1 & 0 & 0 \\
 0 & 0 & -1 & 0 \\
 0 & 0 & 0 & -1 \\
\end{array}
\right)\,,
\end{alignat}
where $\hat{I},\hat{J} = \{\phi^+,\phi^0,\phi^-,\phi^{0*}\}$ (note that the labels above will become clear in the next section). We summarize the eigenvalues in Table \ref{table:weight:vects} and Figure \ref{fig:weights:higgs}.
\begin{table}[!ht]
  \centering
   \begin{tabular}{| c |c | c | c | c| }
      \hline
      Particle Label & $T_{3} = T_{W_3}$ & $T_4 =T_{B}$ & $X_4 = Q$  & $X_4 = Q$ \\
      \hline \hline
      $\phi^+$ & $\frac{1}{2}\,g_1$ & $\frac{1}{2}\,g_2\,y_\phi$ & $\frac{g_1 g_2}{\sqrt{g_1^2 + g_2^2 y_\phi^2}}  y_\phi$ & $e$\\ \hline
      $\phi^0$ & $-\frac{1}{2}\,g_1$ & $\frac{1}{2}\,g_2\,y_\phi$ & $0$ & $0$\\ \hline
      $\phi^-$ & $-\frac{1}{2}\,g_1$ & $-\frac{1}{2}\,g_2\,y_\phi$ & $-\frac{g_1 g_2 }{\sqrt{g_1^2 + g_2^2 y_\phi^2}} y_\phi$ & $-e$\\ \hline
      $\phi^{0*}$ & $\frac{1}{2}\,g_1$ & $-\frac{1}{2}\,g_2\,y_\phi$ & $0$ & $0$ \\ 
      \hline \hline
      $\nu_L$ & $\frac{1}{2}\,g_1$ & $\frac{1}{2}\,g_2\,y_L$ & $\frac{g_1 g_2}{\sqrt{g_1^2 + g_2^2 y_\phi^2}}\frac{y_\phi + y_L}{2}$ &  $0$ \\ \hline
      $e_L$ & $-\frac{1}{2}\,g_1$ & $\frac{1}{2}g_2\,y_L$ & $\frac{g_1 g_2}{\sqrt{g_1^2 + g_2^2 y_\phi^2}}\frac{-y_\phi + y_L}{2}$ & $-e$ \\ \hline
      $e_R$ & $0$ & $\frac{1}{2}\,g_2\,y_{e}$ & $\frac{g_1 g_2}{\sqrt{g_1^2 + g_2^2 y_\phi^2}}\frac{y_{e}}{2}$ & $-e$ \\ \hline
      $\nu_R$ & $0$ & $\frac{1}{2}\,g_2\,y_{\nu}$ &  $\frac{g_1 g_2}{\sqrt{g_1^2 + g_2^2 y_\phi^2}}\frac{y_{\nu}}{2}$ & $0$\\  \hline
      $\nu_L^*$ & $-\frac{1}{2}\,g_1$ & $-\frac{1}{2}\,g_2\,y_L$ &  $-\frac{g_1 g_2}{\sqrt{g_1^2 + g_2^2 y_\phi^2}}\frac{y_\phi + y_L}{2}$ & $0$ \\ \hline
      $e_L^*$ & $\frac{1}{2}\,g_1$ & $-\frac{1}{2}g_2\,y_L$ &  $\frac{g_1 g_2}{\sqrt{g_1^2 + g_2^2 y_\phi^2}}\frac{y_\phi - y_L}{2}$ & $e$\\ \hline
      $e_R^*$ & $0$ & $-\frac{1}{2}\,g_2\,y_{e}$ & $\frac{g_1 g_2}{\sqrt{g_1^2 + g_2^2 y_\phi^2}}\frac{-y_{e}}{2}$ &$e$  \\ \hline
      $\nu_R^*$ & $0$ & $-\frac{1}{2}\,g_2\,y_{\nu}$ &  $\frac{g_1 g_2}{\sqrt{g_1^2 + g_2^2 y_\phi^2}}\frac{-y_\nu}{2}$ & $0$\\ 
      \hline \hline
      $W^+$ & $g_1$ & $0$ &  $\frac{g_1 g_2}{\sqrt{g_1^2 + g_2^2 y_\phi^2}}  y_\phi$ &  $e$\\ \hline
      $W_3$ & $0$ & $0$ & $0$ & $0$ \\ \hline
      $W^-$ & $-g_1$ & $0$ & $-\frac{g_1 g_2}{\sqrt{g_1^2 + g_2^2 y_\phi^2}}  y_\phi$ & $-e$  \\ \hline
      $B$ & $0$ & $0$ & $0$ & $0$  \\      \hline 
   \end{tabular}\caption{Weight vectors for the SM spectrum. In the last column, we set $y_\phi = -y_L = -y_e/2$, and $y_\nu = 0$.\label{table:weight:vects}}
\end{table}

\begin{figure}[h]
\centering
\includegraphics[width=0.49\textwidth]{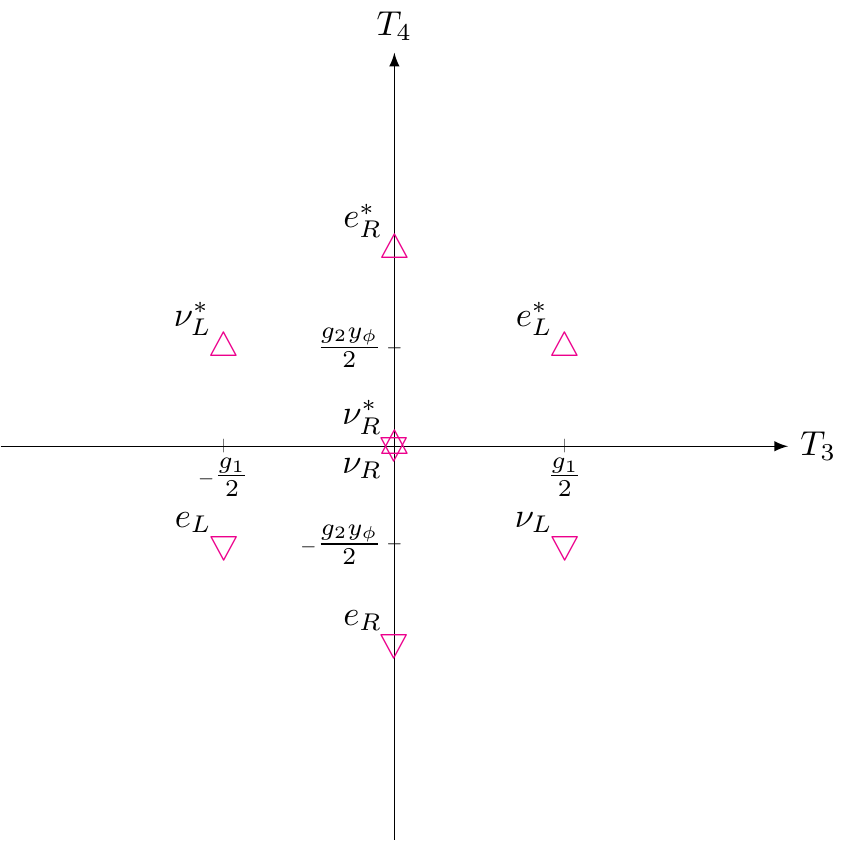} 
\includegraphics[width=0.49\textwidth]{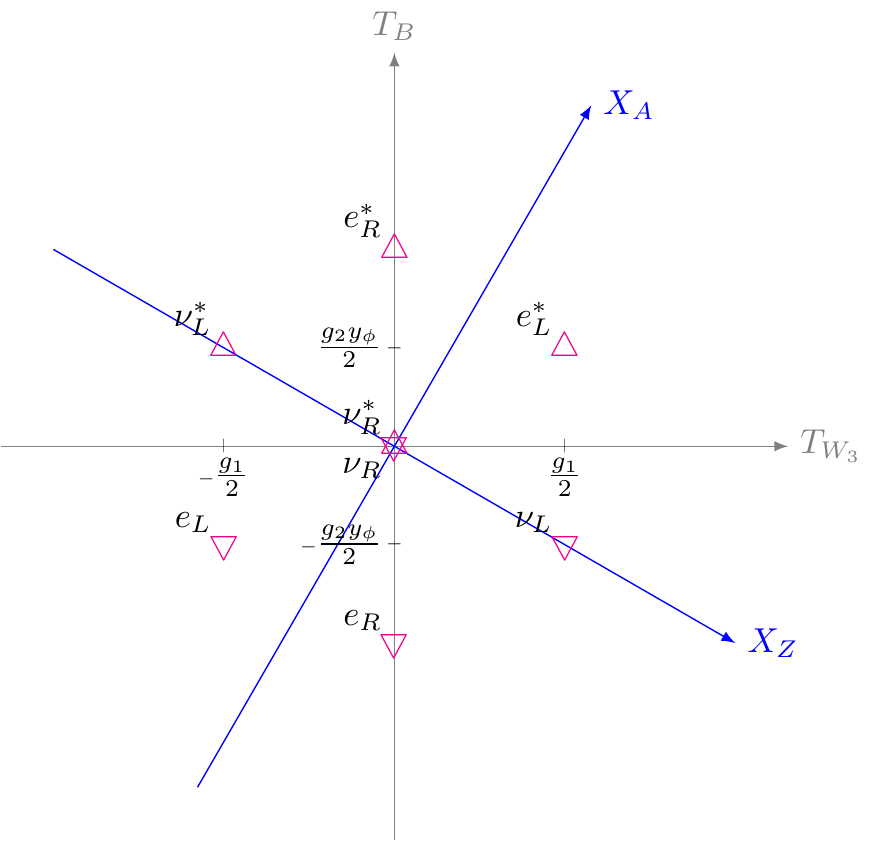}
\caption{The weight diagram for the spin-$1/2$ sector of the Standard Model.\label{fig:weights:fermion}}
\end{figure}

Next, we move onto the spin-$1/2$ sector. The $L$ is a doublet under $SU(2)_L$ and has hypercharge $Y=y_l$, $e_R$ is a singlet under $SU(2)$ and has hypercharge $Y=y_{e}$, and $\nu_R$ is a singlet under $SU(2)_L$ and has hypercharge $Y= y_{\nu}$. Since $L$ is a doublet, we have two extra degrees of freedom which we label as $L = \nu_L, e_L$. Hence, we can construct the reducible representation under which the fermions trasnsform, using $i,j = \{\nu_L, e_L, e_R, \nu_R\}$,
\begin{alignat}{4}\label{eq:sm:fermion:repcomplex}
\tau^f_1 &= \frac{g_1}{2}\left(
\begin{array}{cccc}
 0 & 1 & 0 & 0 \\
 1 & 0 & 0 & 0 \\
 0 & 0 & 0 & 0 \\
 0 & 0 & 0 & 0 \\
\end{array}
\right)\,,&\quad \tau^f_2 &= \frac{g_1}{2} \left(
\begin{array}{cccc}
 0 & -i & 0 & 0 \\
 i & 0 & 0 & 0 \\
 0 & 0 & 0 & 0 \\
 0 & 0 & 0 & 0 \\
\end{array}
\right), \nonumber
\\
\tau^f_3 &= \frac{g_1}{2}\left(
\begin{array}{cccc}
 1 & 0 & 0 & 0 \\
 0 & -1 & 0 & 0 \\
 0 & 0 & 0 & 0 \\
 0 & 0 & 0 & 0 \\
\end{array}
\right),&\quad
\tau^f_4 &= \frac{g_2}{2}  \left(
\begin{array}{cccc}
 y_{L} & 0 & 0 & 0 \\
 0 & y_{L} & 0 & 0 \\
 0 & 0 & y_{e_R} & 0 \\
 0 & 0 & 0 & y_{\nu_R} \\
\end{array}
\right)\,.
\end{alignat}
Since this representation is in a complex basis, and we would like to consider all the real degrees of freedom explicitly, we follow the exact procedure as we did with the scalars (see Appendix~\ref{appendix:repofHiggs}). Notably, we upgrade our representation matrices above to,
\begin{align}\label{eq:sm:fermion:repcomplex:reducible}
T^f_a &= \begin{pmatrix} \tau^f_a & \\
& - (\tau^f_a)^* \end{pmatrix} \,.
\end{align}
where the indexes $i,j = \{ \nu_L, e_L, e_R, \nu_R, \nu_L^*, e_L^*, e_R^*, \nu_R^*\}$ now run over the conjugate representation as well. Note that the asterisks are just part of the label names, and only represent the basis that the conjugate representation of the group $G$ is defined over. Further, this basis is still complex, but we are simply a change of basis away from the real degrees of freedom. We summarize the eigenvalues in Table \ref{table:weight:vects} and Figure \ref{fig:weights:fermion}.

\begin{figure}[!ht]
\centering
\includegraphics[width=0.49\textwidth]{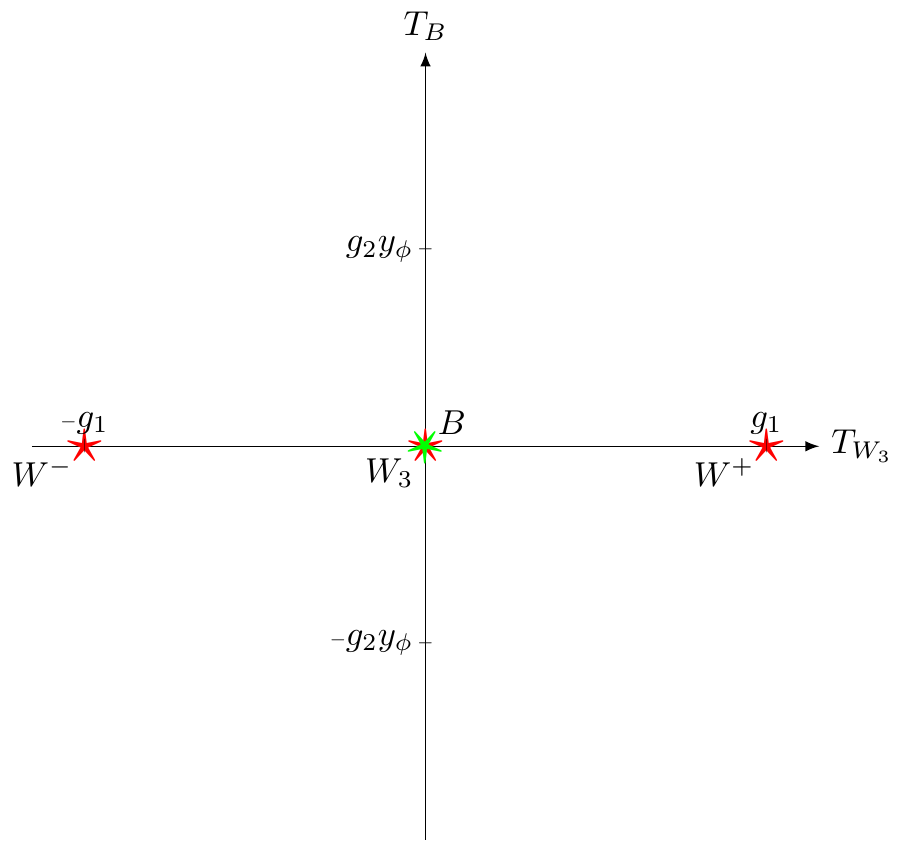} 
\includegraphics[width=0.49\textwidth]{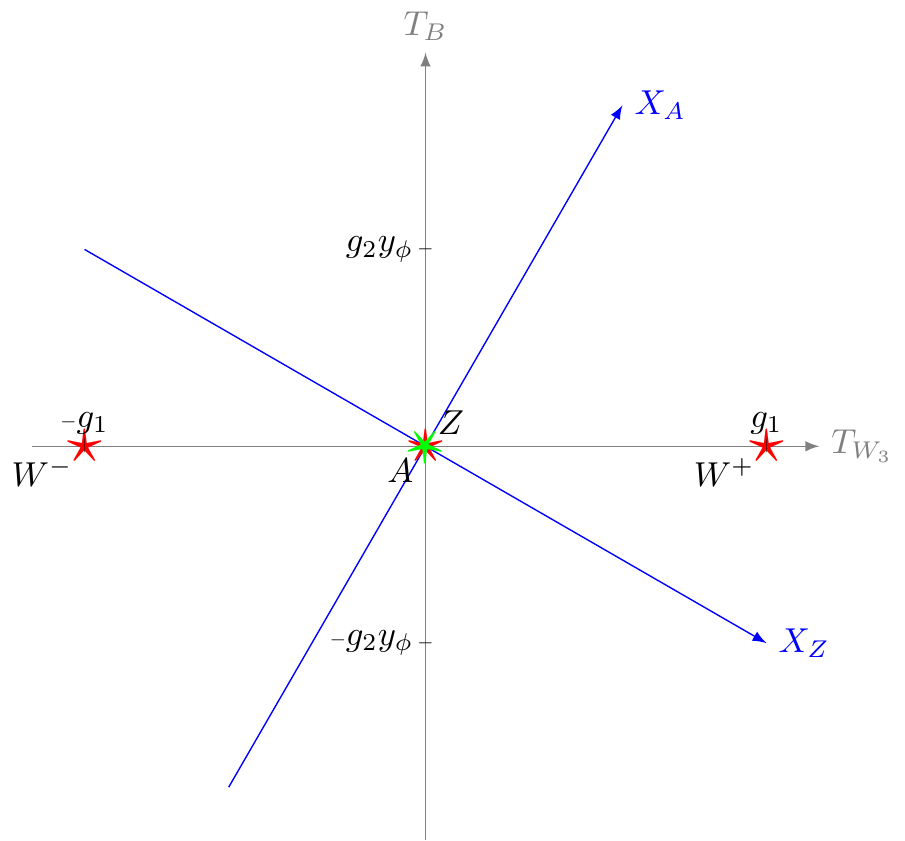}
\caption{The weight diagram of the spin-$1$ particles of the Standard Model.\label{fig:weights:bosons}}
\end{figure}

The spin-$1$ particles transform under the adjoint representation. Since this is defined through the group $G$, we have no freedom to select charges. Using eq.~\eqref{eq:compute:structureconstant}, we can use the representations $R=s,f$, or simply the definitions of $SU(2)\times U(1)$, to first get the structure constant in the UV,
\begin{equation}
   \fuv_{\hat{a}\hat{b}\hat{c}} = \left(
\begin{array}{cccc}
 \{0,0,0,0\} & \{0,0,i g_1,0\} & \{0,-i g_1,0,0\} & \{0,0,0,0\} \\
 \{0,0,-i g_1,0\} & \{0,0,0,0\} & \{i g_1,0,0,0\} & \{0,0,0,0\} \\
 \{0,i g_1,0,0\} & \{-i g_1,0,0,0\} & \{0,0,0,0\} & \{0,0,0,0\} \\
 \{0,0,0,0\} & \{0,0,0,0\} & \{0,0,0,0\} & \{0,0,0,0\} \\
\end{array}
\right)
\end{equation}
which summarizes that the only interactions in the UV are permutations of $\{W_1,W_2,W_3\}$. The $B$ does not interact with any of the other bosons. Then, we can determine the adjoint representaion via $f_{\hat{a}\hat{b}\hat{c}}=-(T^\text{ad}_{\hat{a}})_{\hat{b}\hat{c}}$,
\begin{align}
   T^\text{ad}_1 = g_1 \left(
\begin{array}{cccc}
 0 & 0 & 0 & 0 \\
 0 & 0 & -i  & 0 \\
 0 & i  & 0 & 0 \\
 0 & 0 & 0 & 0 \\
\end{array}
\right),T^\text{ad}_2 = g_1\left(
\begin{array}{cccc}
 0 & 0 & i & 0 \\
 0 & 0 & 0 & 0 \\
 -i  & 0 & 0 & 0 \\
 0 & 0 & 0 & 0 \\
\end{array}\right)T^\text{ad}_3 = g_1\left(
\begin{array}{cccc}
 0 & -i & 0 & 0 \\
 i & 0 & 0 & 0 \\
 0 & 0 & 0 & 0 \\
 0 & 0 & 0 & 0 \\
\end{array}
\right),T^\text{ad}_4 = g_2\left(
\begin{array}{cccc}
 0 & 0 & 0 & 0 \\
 0 & 0 & 0 & 0 \\
 0 & 0 & 0 & 0 \\
 0 & 0 & 0 & 0 \\
\end{array}
\right)
\end{align}
The usual strategy is to diagonalize these, however, instead of changing our basis explicitly, we can determine their values under measurement of $T_3^{\text{ad}}$ via commutation, which is how the adjoint representation acts. Defining $T_{W^\pm}^{R} \equiv \frac{1}{\sqrt{2}}(T_1^{R} \pm i T_2^{R})$,
\begin{align}
    T_3^{R}|T_{W^+}^{R}\ra &=  |[T_3^{R},T_{W^+}^{R}]\ra = g_1 |T_{W^+}^{R}\ra \,,\nonumber \\
    T_3^{R}|T_{W^-}^{R}\ra &=  |[T_3^{R},T_{W^-}^{R}]\ra = -g_1 |T_{W^-}^{R}\ra \,,\nonumber \\
    T_3^{R}|T_3^{R}\ra &=  |[T_3^{R},T_3^{R}]\ra = 0 \,,\nonumber \\
    T_3^{R}|T_4^{R}\ra &=  |[T_3^{R},T_4^{R}]\ra = 0 \,,
\end{align}
and $T_B^R|T_a^R\ra = 0$ for all $a$, where the same results carry though for any representation $R$. Its worth mentioning here that the $W^\pm$ labels are purely associated with the fact that adjoint representation for $SU(2)$ is the spin-$1$ representation, and so the states $T_{W^-}$,$T_{3}$, and $T_{W^+}$ form a triplet. Indeed, $T_{W^+}|T_{W^-}\ra = |[T_{W^+},T_{W^-}]\ra = g_1|T_{3}\ra$, and $T_{W^+}|T_{3}\ra = |[T_{W^+},T_{3}]\ra = -g_1|T_{W^+}\ra$. So $T_{W^+}$ raises the states in units of $g_1$. We have not mentioned anything about $U(1)_{EM}$. We summarize the spin-$1$ spectrum in Figure \ref{fig:weights:bosons}.
The summary of Table \ref{table:weight:vects} for all particles is shown in 
\begin{figure}[h]
\centering
\includegraphics[width=0.49\textwidth]{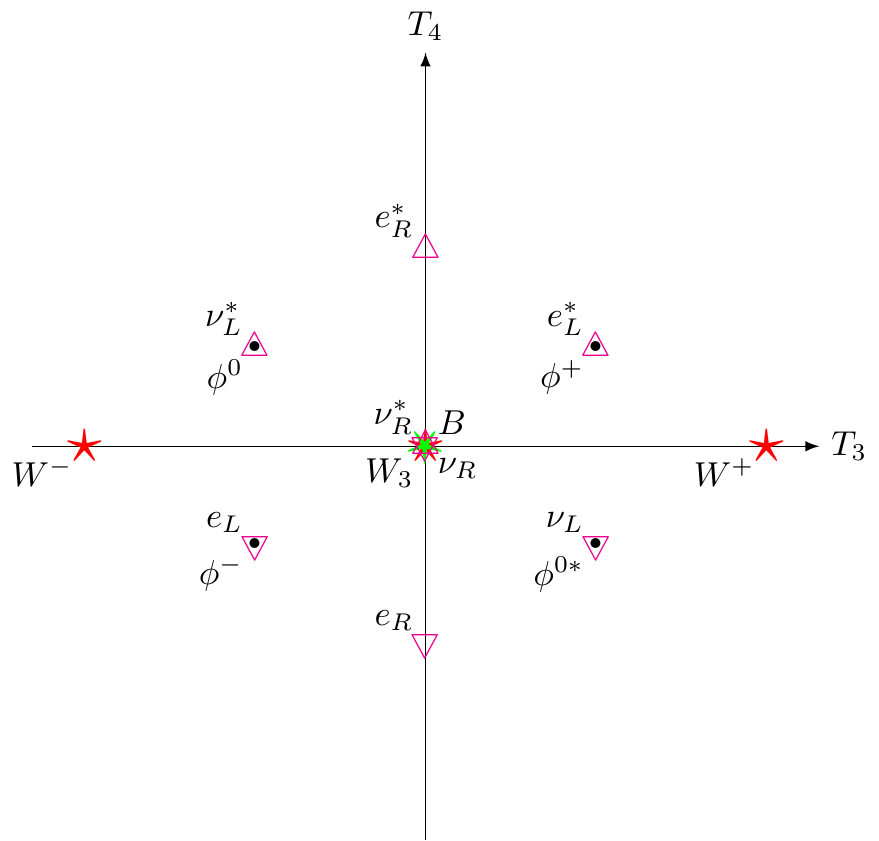} 
\includegraphics[width=0.49\textwidth]{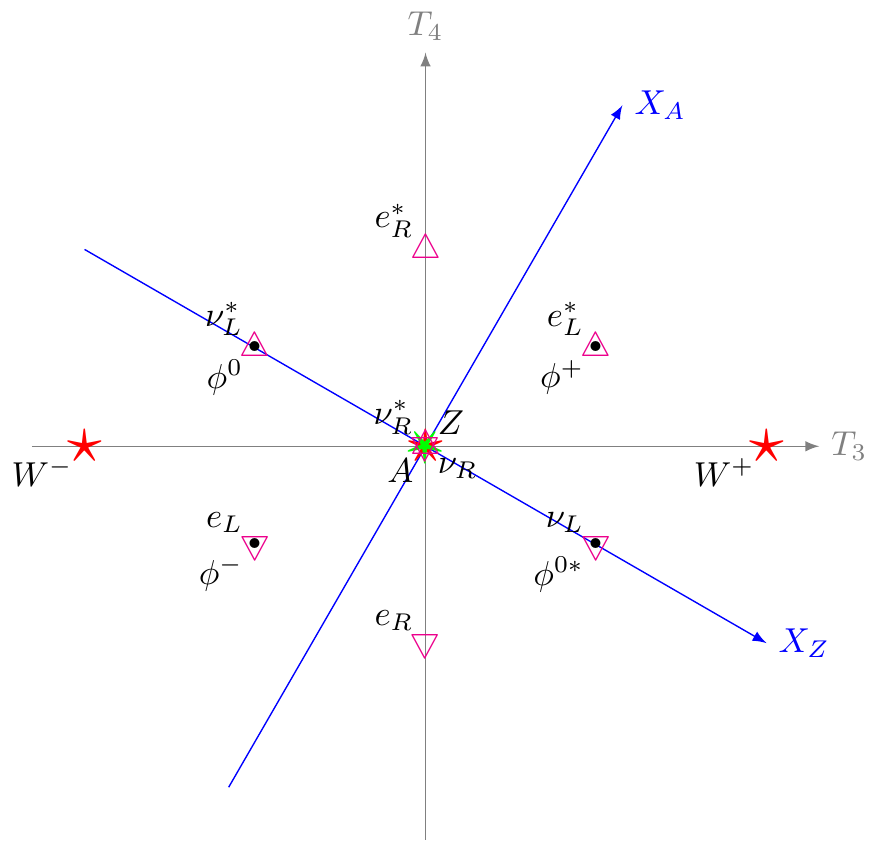}
\caption{The weight diagram for the spin-$0$, $1/2$, and $1$ particles of the Standard Model.\label{fig:weights:all}}
\end{figure}

The Yukawa couplings $Y_{\hat{I}\hat{j}\hat{k}}$ must satisfy eq.~\eqref{eq:yukawa:constraint}. 
Setting $y_\phi = -y_L = -y_e/2$, and $y_\nu = 0$ allows us to determine the non-zero Yukawa couplings and gauge invariant mass terms.

\subsection{IR}

The first thing to address it the presence of a massless particle $a = A$ in the IR, known as the photon. This means we must have a vector $\vec{v}$ that satisfies eq.~\eqref{eq:massless:spin:1}, notably, $X_A^s \vec{v} = \c{O}_A^{\;\; \hat{a}}T_{\hat{a}}^s \vec{v} = \vec{0}$. In other words, $\vec{v}$ must be an eigenvector of $X_{A}^s$ with eigenvalue $0$.
Posed in this way, we can use the weight diagrams in Figure~\ref{fig:weights:higgs}, which represent the eigenvalues of the Cartan subalgebra of $\mathfrak{g}$, in the representation of the scalars. 
Thus, our only hope of getting a zero eigenvalue is to change the basis of our algebra.
The Weinberg angle $\theta_w$ \cite{Glashow:1961tr} was introduced to do just that,
\begin{equation}\label{eq:rotation:UVIR:SM}
  \c{O}^{\hat{a}}_{\;\; \u{a}}=  \left(
\begin{array}{cccc}
 1 & 0 & 0 & 0 \\
 0 & 1 & 0 & 0 \\
 0 & 0 & \cos \theta_w  & \sin \theta_w  \\
 0 & 0 & -\sin \theta_w  & \cos \theta_w  \\
\end{array}
\right)\,.
\end{equation}
Note that nothing has required us to make rotations in $W_1$ and $W_2$. Now, one can easily solve for $\theta_w$,
\begin{align}\label{eq:weinberg:angle:solution}
    \tan \theta_w =\frac{g_2y_\phi}{g_1} \, \Rightarrow \cos\theta_w = \frac{g_1}{\sqrt{g_1^2 + g_2^2 y_\phi^2}} \,, \quad \text{and} \quad \sin\theta_w = \frac{ g_2 y_\phi}{\sqrt{g_1^2 + g_2^2 y_\phi^2}}\,.
\end{align}
To make contact with usual descriptions, we can write the generators in the IR by implementing the rotation,
\begin{align}\label{eq:sm:rotated:gens}
    X_1^R = T_1^R\,,\quad X_2^R = T_2^R\,,\quad X_3^R = c_w T_3^R - s_w T_4^R \,,\quad X_4^R = s_w T_3^R + c_w T_4^R\,.
\end{align}
The alignment of $X_3^s$ and $X_4^s$ is shown in Figure~\ref{fig:weights:higgs}.
Using the identification $\u{a},\u{b},\u{c} \in \{1,2,3,4\} = \{W^1,W^2,Z,A\}$, it is clear that the `un-broken' generator $X^s_4 = X^s_A = Q^s$.  Putting back in the coupling constants, we have the familiar,
\begin{align}
    Q^s &= \frac{g_1g_2y_\phi}{\sqrt{g_1^2 + g_2^2y_\phi^2}} \left( \tilde{T}_3 +\tilde{T}_4 \right) = e\left( \tilde{T}_3 +\tilde{T}_4 \right) \,,
\end{align}
where we have identified the $e =  \frac{g_1g_2y_\phi}{\sqrt{g_1^2 + g_2^2y_\phi^2}}$.
Explicitly,
\begin{align}
    X^s_4 =\left(
\begin{array}{cccc}
 e & 0 & 0 & 0 \\
 0 & 0 & 0 & 0 \\
 0 & 0 & -e & 0 \\
 0 & 0 & 0 & 0\\
\end{array}
\right) \,,
\end{align}
which now makes the notation $\hat{I},\hat{J} = \{\phi^+,\phi^0,\phi^-,\phi^{0*}\}$ hopefully clear.
Since there is only one generator for which $X_{a}^s\vec{v} = \vec{0}$, our subalgebra $\mathfrak{h} = Q$ and generates a $U(1)$ group, referred to as $U(1)_{EM}$.

With the rotations in eq.~\eqref{eq:sm:rotated:gens}, we can re-measure IR spin-$1$ particles with $X_4$. Since we have kept $X_1 = T_1$ and $X_2 = T_2$, then $X_{W^\pm}^{R} =T_{W^\pm}^{R} \equiv \frac{1}{\sqrt{2}}(T_1^{R} \pm i T_2^{R})$,
\begin{align}
    X_4^{R}|X_{W^+}^{R}\ra &=  |[X_4^{R},X_{W^+}^{R}]\ra = e |X_{W^+}^{R}\ra \,,\nonumber \\
    X_4^{R}|X_{W^-}^{R}\ra &=  |[X_4^{R},X_{W^-}^{R}]\ra = -e |X_{W^-}^{R}\ra \,,\nonumber \\
    X_4^{R}|X_3^{R}\ra &=  |[X_4^{R},X_
    3^{R}]\ra = 0 \,,\nonumber \\
    X_4^{R}|X_4^{R}\ra &=  |[X_4^{R},X_4^{R}]\ra = 0 \,,
\end{align}
which tells us that the $Z$ and photon $A$ are neutral, and the $W^\pm$ have charges $\pm e$ under $U(1)_{EM}$. Thus, the $\pm$ labels on $W^\pm$ serve a dual purpose, of labeling the charges under $T_3$ as well as $X_4$.

Further, with $\theta_w$ and $\c{O}$, we can easily compute the couplings between all massless and massive gauge bosons (or the `structure constant' in the IR). Using $\u{a},\u{b},\u{c} \in \{1,2,3,4\} = \{W^1,W^2,Z,A\}$,
\begin{align}
h_{\u{a}\u{b}\u{c}} &= \fuv_{\hat{a}\hat{b}\hat{c}}  \c{O}^{\hat{a}}_{\;\; \u{a}} \c{O}^{\hat{b}}_{\;\; \u{b}} \c{O}^{\hat{c}}_{\;\; \u{c}} \nonumber \\
  &= \left(
\begin{array}{cccc}
 \{0,0,0,0\} & \{0,0,i g_1c_w,i g_1s_w\} & \{0,-i g_1c_w,0,0\} & \{0,-i g_1s_w,0,0\} \\
 \{0,0,-i g_1c_w,-i g_1s_w\} & \{0,0,0,0\} & \{i g_1c_w,0,0,0\} & \{i g_1s_w,0,0,0\} \\
 \{0,i g_1c_w,0,0\} & \{-i g_1c_w,0,0,0\} & \{0,0,0,0\} & \{0,0,0,0\} \\
 \{0,i g_1s_w,0,0\} & \{-i g_1s_w,0,0,0\} & \{0,0,0,0\} & \{0,0,0,0\} \\
\end{array}
\right)
\end{align}
From this, it is clear that the $Z = 3$ only interacts with $W_1 =1$ and $W_2 =2$, and similarly, the  photon $A = 4$ only interacts with $W_1 =1$ and $W_2 =2$. Identical bosons do not interact, as seen by the diagonal of zeros. The $W_1 =1$ can interact with the $W_2$, $Z$ and $A$. Not only have we found all the allowed interactions (for massless and massive bosons), but we have also listed all of their couplings!

Now, let us consider the eigenvector $\vec{v}$. Note that we have yet to say anything about $\vec{v}$ itself. We have only demanded the existence of a non-trivial solution to $X_A^s \vec{v} = \vec{0}$, which was a statement about eigenvalues. In the current reducible complex basis, the most general $\vec{v}$ takes the form $\vec{v} = (0,v_3,0,v_4)$. 
From Appendix~\ref{appendix:repofHiggs}, in the real basis,
\begin{align}
    \vec{v} \rightarrow A^{-1}\vec{v} = \frac{1}{\sqrt{2}}\begin{pmatrix}0\\ v_3 + v_4\\0\\-v_3i + v_4i\end{pmatrix}
\end{align}
which tells us that $v_3 = v_4 = v\in \mathbb{R}$. So, in the complex reducible basis $\vec{v} = v(0,1,0,1)$ and in the real basis $\vec{v} = v\sqrt{2}(0,1,0,0)$. In the real basis, it is clear that there is an $SO(3)$ symmetry of $\vec{v}$, known as the `custodial $SU(2)$' \cite{Sikivie:1980hm}.

With the structure of $\vec{v}$ fixed, we can compute the masses of the spin-$1$ particles. Of course, the answer must be basis independent, and using either the complex reducible basis or real basis, with $m^2_{\u{a}\u{b}} =\vec{v}^TX_{\u{a}}^{sT} X_{\u{b}}^s \vec{v}$
\begin{align}
    m^2 &= \frac{v^2}{2}\left(
\begin{array}{cccc}
  g_1^2 & 0 & 0 & 0 \\
 0 &  g_1^2 & 0 & 0 \\
 0 & 0 &  g_1^2  + g_2^2\, y_\phi^2 & 0 \\
 0 & 0 & 0 & 0 \\
\end{array}
\right) = \left(
\begin{array}{cccc}
  m_{W_1}^2 & 0 & 0 & 0 \\
 0 &  m_{W_2}^2 & 0 & 0 \\
 0 & 0 & m_{Z}^2 & 0 \\
 0 & 0 & 0 & m_{A}^2 \\
\end{array}
\right) \,.
\end{align}

Next, we can consider the masses of the fermions. As mentioned in eq.~\eqref{eq:solution4}, the masses are given by
$ m_{\u{k}} \delta^{\;\u{i}}_{\u{k}}=  [\Omega^T (v^{\hat{I}}Y_{\hat{I}\hat{j}\hat{k}})\Omega]_{\u{k}}^{\;\u{i}}$. In the complex basis with $\hat{I}= \{\phi^+,\phi^0,\phi^-,\phi^{0*}\}$ and $\vec{v} =v(0,1,0,1)$, the non-zero contributors to this are
\begin{align}
    v^{\hat{I}}Y_{\hat{I}\hat{j}\hat{k}} + \tilde{m}&= v \left(Y_{\phi^0\hat{j}\hat{k}} + Y_{\phi^{0*}\hat{j}\hat{k}}\right) \,,
\end{align}
which, restricting to the electron degrees of freedom, yields the singular values,
\begin{align}
    \Omega^T  v^{\hat{I}}Y_{\hat{I}\hat{j}\hat{k}} \Omega = v |Y_e| \text{diag} (1,1,1,1)\,.
\end{align}
So the masses are degenerate with $m_e = v |Y_e|$. In addition, the most general mass term for the neutrinos can be similarly determined, and gives the well known see-saw mechanism.

\section{Conclusion}

We have shown a completely on-shell version of spontaneous symmetry breaking, and how it can generate pattern of masses for spins $1/2$ and $1$ particles, as well as interactions.  The two main ingredients were unitary in the form of consistent factorization, as well as UV IR consistency. We were able to land on many of the results one would arrive at from QFT. As a final remark, we showed how the SM fits into this formalism. 

This continues the efforts of connecting modern on-shell methods to the real world, and a natural extension of this work would be to consider supersymmetry (SUSY). It would be interesting to see the constraints of consistent factorization on SUSY interactions, as well as the on-shell version of SUSY breaking. 

\acknowledgments
I thank Nima Arkani-Hamed and Akashay Yelleshpur for guidance at all stages of this project. I would also like to thank Sebastian Mizera for reviewing a draft and providing comments. Lastly, I thank Nicholas Haubrich for interesting discussions related to the Standard Model while approaching this problem.

\appendix

\section{Conventions}

We use the metric signature $(+,-,-,-)$. The $SL(2,C)$ spinor indices $\alpha$, $\beta$, $\dot\alpha$ and $\dot\beta$ are raised and lowered using
\begin{align}
\epsilon_{\alpha\beta}= - \epsilon^{\alpha\beta} = \epsilon_{\dot\alpha\dot\beta} = - \epsilon^{\dot{\alpha}\dot{\beta}}  = 
\begin{pmatrix} 0 & -1 \\
1 & 0 
\end{pmatrix}   \,.
\end{align}
The $SU(2)$ little group indices $I,J$ are raised and lowered using
\begin{align}
  \epsilon_{IJ} = - \epsilon^{IJ} = \begin{pmatrix} 0 & -1 \\
1 & 0 
\end{pmatrix}   \,.
\end{align}
We also use the Weyl/Chiral representation of the   Dirac algebra,
\begin{align}
  \gamma^\mu = \begin{pmatrix}
    0 & (\sigma^\mu)_{\alpha\dot\beta} \\
    (\bar{\sigma}^\mu)^{\dot\alpha\beta} & 0
  \end{pmatrix}\,,
\end{align}
with $\sigma^\mu = (1,\sigma^i)$ and $\bar\sigma^\mu = (1,-\sigma^i)$ with $\sigma^{1,2,3}$. 

\section{Kinematics\label{sec:kinematics}}
We can rewrite $p_\mu$ as a Weyl bi-spinor via,
\begin{align}
  p_\mu \gamma^\mu &= \begin{pmatrix}
                    0 & p_\mu(\sigma^\mu)_{\alpha\dot\beta} \\
                    p_\mu(\bar{\sigma}^\mu)^{\dot\alpha\beta} & 0
                  \end{pmatrix} =\begin{pmatrix}
                                  0 & p_{\alpha\dot\beta} \\
                                  p^{\dot\alpha\beta} & 0
                                \end{pmatrix} \,,
\end{align}
which implements the isomorphism between the Lorentz group $SO(3,1)$ and $SL(2,C)\times SL(2,C)$. Explicitly,
\begin{align}
  p_{\alpha\dot\beta} = \begin{pmatrix} 
                       p_0 +p_3 & p_1 - i p_2 \\
                          p_1 + i p_2 & p_0 - p_3
                        \end{pmatrix} \,, \quad \text{and}\quad p^{\dot\alpha\beta} &= \begin{pmatrix} 
                          p_0 -p_3 & -p_1 + i p_2 \\
                          -p_1 - i p_2 & p_0 + p_3
                        \end{pmatrix} \,.
\end{align}
Raising and lowering of the $SL(2,C)$ indices are defined by $(\bar\sigma^\mu)^{\dot\alpha\alpha}=\epsilon^{\alpha\beta}\epsilon^{\dot\alpha\dot\beta}(\sigma^\mu)_{\beta\dot\beta}$. For example, contracting with $p_\mu$ gives $p^{\dot\alpha\alpha} = \epsilon^{\alpha\beta}\epsilon^{\dot\alpha\dot\beta}p_{\beta\dot\beta}$.
Lastly, we note that 
\begin{align}
  \det(p_{\alpha\dot\alpha}) = \det(p^{\dot\alpha\alpha}) =p^2 = m^2\,,
\end{align}
which can be seen explicitly from the matrices above, or using the fact that $(\sigma^\mu)_{\alpha\dot\alpha}(\bar\sigma^\nu)^{\dot\alpha\alpha} = 2 \eta^{\mu\nu}$, to simplify $\det(p_{\alpha\dot\alpha}) = \frac{1}{2} \epsilon^{\dot\alpha\dot\beta}\epsilon^{\alpha\beta}p_{\alpha\dot\alpha}p_{\beta\dot\beta} = \frac{1}{2}p_{\alpha\dot\alpha}p^{\dot\alpha\alpha} = p^2$. Hence, we see a clear distinction between massless and massive particles, which we will discuss next.

\subsection{Massless Particles and Helicity-Spinors}
For massless particles, $\det(p_{\alpha\dot\alpha}) =\det(p^{\dot\alpha\alpha})= 0$, and so the matrices $p_{\alpha\dot\alpha}$ and $p^{\dot\alpha\alpha}$ have rank $1$. Hence, we can write it as a direct product of two spinors, 
\begin{align}
  p_{\alpha\dot\alpha} &= \lambda_\alpha \tilde\lambda_{\dot\alpha} \equiv |\lambda\rangle_\alpha[\tilde\lambda|_{\dot\alpha}\, \qquad \text{and}\qquad p^{\dot\alpha\alpha} = \tilde\lambda^{\dot\alpha}\lambda^\alpha \equiv |\tilde\lambda]^{\dot\alpha}\langle\lambda|^\alpha \,,
\end{align}
where we have introduced the Dirac bra-ket notation. Keeping in mind $p^{\dot\alpha\alpha} = \epsilon^{\alpha\beta}\epsilon^{\dot\alpha\dot\beta}p_{\beta\dot\beta}$, it is easy to see that the spinors are related in the way we expect,
\begin{align}
  \lambda^\alpha = \epsilon^{\alpha\beta}\lambda_\beta \qquad \text{and} \qquad \tilde\lambda^{\dot\alpha} = \epsilon^{\dot\alpha\dot\beta} \tilde\lambda_{\dot\beta}\,,
\end{align}
or equivalently
\begin{align}
  \la\lambda|^\alpha = \epsilon^{\alpha\beta}|\lambda\ra_\beta \qquad \text{and} \qquad |\tilde\lambda]^{\dot\alpha} = \epsilon^{\dot\alpha\dot\beta} [\tilde\lambda|_{\dot\beta}\,.
\end{align}
For real momenta $(p_{\alpha\dot\alpha})^* = (p_{\alpha\dot\alpha})^T$, which implies $(\lambda_\alpha)^* = \tilde\lambda_{\dot\alpha}$.

As mentioned before, spinors are especially useful because they transform under both the Lorentz group and little group. The action of the little group can be observed by the scaling
\begin{align}
  \lambda_\alpha \rightarrow \lambda^\prime_\alpha = w \lambda_\alpha \qquad \text{and} \qquad \tilde\lambda_{\dot\alpha} \rightarrow \tilde\lambda^\prime_{\dot\alpha} = w^{-1}\tilde\lambda_{\dot\alpha}\,,
\end{align}
which leaves $p_{\alpha\dot\alpha}$ invariant. For real momenta, to preserve $(\lambda_\alpha)^* = \tilde\lambda_{\dot\alpha}$ under the transformation, $(\lambda^\prime_\alpha)^* = \tilde\lambda^\prime_{\dot\alpha}$, we must have $w^{-1} = w^*$ i.e. $w$ must be an element of $U(1)$.

The massless Dirac equation follows immediately
\begin{align}
  p_{\alpha\dot\alpha}\tilde\lambda^{\dot\alpha} = 0 \quad \text{and} \quad p^{\dot\alpha\alpha}\lambda_\alpha = 0 \,,
\end{align}
since $\lambda^\alpha\lambda_\alpha = \tilde\lambda_{\dot\alpha} \tilde\lambda^{\dot\alpha} = 0$.

The above conditions are enough to find explicit representations for the spinors. Picking a general momentum $p^\mu = (E,|\vec{p}|\sin\theta\cos\phi, |\vec{p}|\sin\theta\cos\phi, |\vec{p}|\cos\theta)$, yields
\begin{align}\label{eq:massless:spinors:explicit}
  \lambda_\alpha = \sqrt{2E}\begin{pmatrix}
                              c \\ s
                            \end{pmatrix}
                            \quad
                            \text{and}
                            \quad
  \tilde\lambda^{\dot\alpha} = \sqrt{2E}\begin{pmatrix}
                              s^*\\ -c
                            \end{pmatrix}\,,                          
\end{align}
with
$c = \cos\theta/2$ and $s = e^{i\phi}\sin\theta/2$.

Finally, we note that these spinors are also eignstates of the helicity operator $\Sigma = S.\vec{p}/|\vec{p}|$ where spin $S_i = \frac{1}{2}\epsilon_{ijk}S^{jk}$ is defined by the spin matrix $S^{\mu\nu} = \frac{i}{4}[\gamma^\mu,\gamma^\nu]$,
\begin{align}\label{eq:helicity:def:spinor}
  \Sigma \lambda_\alpha = +\frac{1}{2}\lambda_\alpha \quad \text{and} \quad \Sigma \tilde\lambda^{\dot\alpha} = -\frac{1}{2} \tilde\lambda^{\dot\alpha}\,.
\end{align}

\subsection{Massive Particles and Spin-Spinors}
For massive particles, $\det(p_{\alpha\dot\alpha}) =\det(p^{\dot\alpha\alpha}) = m^2$, and so the matrices $p_{\alpha\dot\alpha}$ and $p^{\dot\alpha\alpha}$ have rank $2$, and can be written as a sum of two rank $1$ matrices,
\begin{align}\label{eq:massive:momentum:spinor}
  p_{\alpha\dot\alpha}= \b{\lambda}_\alpha^{\;\; I} \tilde{\b{\lambda}}_{\dot\alpha I} \equiv |\b{\lambda}\ra_\alpha^{\;\;I} [\tilde{\b{\lambda}}|_{\dot\alpha I} \qquad \text{and} \qquad p^{\dot\alpha\alpha}= \tilde{\b{\lambda}}^{\dot\alpha}_{\;\; I} \b{\lambda}^{\alpha I} \equiv |\tilde{\b{\lambda}}]^{\dot\alpha}_{\;\;I}\la\b{\lambda}|^{\alpha I} 
\end{align}
where $I = 1,2$, and again, we have introduced the Dirac bra-ket notation.  As in the massless case, the spinors are related by  
\begin{alignat}{4}
  \b{\lambda}^{\alpha I} &= \epsilon^{\alpha\beta}\b{\lambda}_\beta^{\;\; I} \qquad &&\text{and} \qquad \tilde{\b{\lambda}}^{\dot\alpha I} &&= \epsilon^{\dot\alpha\dot\beta} \tilde{\b{\lambda}}_{\dot\beta}^{\;\; I}\,,\quad &\text{or}\\
  \la\b{\lambda}|^{\alpha I} &= \epsilon^{\alpha\beta}|\b{\lambda}\ra_\beta^{\;\;I} \qquad &&\text{and} \qquad |\tilde{\b{\lambda}}]^{\dot\alpha I} &&= \epsilon^{\dot\alpha\dot\beta} [\tilde{\b{\lambda}}|_{\dot\beta}^{\;\;I}\,.&
\end{alignat}
Again, for real momenta, $(p_{\alpha\dot\alpha})^* = (p_{\alpha\dot\alpha})^T$, which implies $(\b{\lambda}_\alpha^{\;\;I})^* = (\tilde{\b{\lambda}}_{\dot\alpha I})^T$.

We can see the action of the little group by the $SL(2)$ transformations $W$,
\begin{align}
  \b{\lambda}_\alpha^{\;\;I} \rightarrow \b{\lambda}_\alpha^{\;\;I^\prime} = W_{I^\prime}^{\;\;I} \b{\lambda}_\alpha^{\;\;I^\prime} \qquad \text{and} \qquad \tilde{\b{\lambda}}_{\dot\alpha I} \rightarrow \tilde{\b{\lambda}}_{\dot\alpha I^\prime} = (W^{-1})_I^{\;\;I^\prime} \tilde{\b{\lambda}}_{\dot\alpha I^\prime} \,.
\end{align}
For real momenta, to preserve $(\b{\lambda}_\alpha^{\;\;I})^* = \tilde{\b{\lambda}}_{\dot\alpha I}$ under the transformation,$(\b{\lambda}_\alpha^{\;\;I^\prime})^* = \tilde{\b{\lambda}}_{\dot\alpha I^\prime}$, we must have $W^{-1} = W^\dagger$, i.e. $W$ must be an element of $SU(2)$. Hence, the little group indices $I,J$ can be raised and lowered as,
\begin{alignat}{5}
  \b{\lambda}^{\alpha I} &= \epsilon^{IJ}\b{\lambda}^\alpha_{\;\; J} \qquad &&\text{and} \qquad \tilde{\b{\lambda}}^{\dot\alpha I} &&= \epsilon^{IJ} \tilde{\b{\lambda}}^{\dot\alpha}_{\;\; J} \,,\qquad &\text{or}\,\\
  \la\b{\lambda}|^{\alpha I} &= \epsilon^{IJ}\la\b{\lambda}|^\alpha_{\;\; J} \qquad &&\text{and} \qquad |\tilde{\b{\lambda}}]^{\dot\alpha I} &&= \epsilon^{IJ} |\tilde{\b{\lambda}}]^{\dot\alpha}_{\;\; J},.  &
\end{alignat}
This allows us to express the momentum as
\begin{align}
  p_{\alpha\dot\alpha}= \epsilon_{IJ}\b{\lambda}_\alpha^{\;\; I} \tilde{\b{\lambda}}_{\dot\alpha}^{\;\;J} \equiv \epsilon_{IJ}|\b{\lambda}\ra_\alpha^{\;\;I} [\tilde{\b{\lambda}}|_{\dot\alpha}^{\;\;J} \quad \text{and} \quad p^{\dot\alpha\alpha}= \epsilon_{IJ}\tilde{\b{\lambda}}^{\dot\alpha J} \b{\lambda}^{\alpha I} \equiv \epsilon_{IJ}|\tilde{\b{\lambda}}]^{\dot\alpha J}\la\b{\lambda}|^{\alpha I} 
\end{align}

Note that $\det(p_{\alpha\dot\alpha}) = \det(\b{\lambda}_\alpha^{\;\; I})\det(\tilde{\b{\lambda}}_{\dot\alpha I}) = m^2$, which, without loss of generality, can be used to set $\det(\b{\lambda}_\alpha^{\;\; I})=\det(\tilde{\b{\lambda}}_{\dot\alpha I}) = m$. This choice imposes the following identities,
\begin{align}\label{eq:spin-spinor-contract}
  \tilde{\b\lambda}_{\dot\alpha I}\tilde{\b\lambda}^{\dot\alpha}_{\;\;J} = -m\epsilon_{IJ} \quad \text{and} \quad \b{\lambda}^{\alpha I} \b{\lambda}_{\alpha}^{\;\; J} = -m \epsilon^{IJ}\,,
\end{align}
which are necessary to maintain the condition $m^2 = \det(p_{\alpha\dot\alpha})$. This can be made clearer by noting that $\det(p_{\alpha\dot\alpha}) = \frac{1}{2} \epsilon^{\alpha\beta}\epsilon^{\dot\alpha\dot\beta}p_{\alpha\dot\alpha}p_{\beta\dot\beta}$, and after substituting the expansions of momentum in terms of spin-spinors, $m^2 = \frac{1}{2}\b{\lambda}^{\alpha J} \b{\lambda}_\alpha^{\;\; I} \epsilon^{\dot\alpha\dot\beta} \tilde{\b\lambda}_{\dot\alpha I} \tilde{\b\lambda}_{\dot\beta J} = m \det(\tilde{\b\lambda}_{\dot\alpha I})$, where demanding the last equality imposes the identity above.

The above definitions yield the Dirac equation,
\begin{align}
  p_{\alpha\dot\alpha}\tilde{\b{\lambda}}^{\dot\alpha}_{\;\;I} = m \b{\lambda}_{\alpha I} \quad \text{and} \quad p^{\dot\alpha\alpha}\b{\lambda}_\alpha^{\;\;I} = m \tilde{\b\lambda}^{\dot\alpha I},
\end{align}
which is easily seen by, for example, $p_{\alpha\dot\alpha}\tilde{\b{\lambda}}^{\dot\alpha}_{\;\;J} =  \b{\lambda}_\alpha^{\;\; I} \tilde{\b{\lambda}}_{\dot\alpha I}\tilde{\b{\lambda}}^{\dot\alpha}_{\;\;J}$ and then applying eq.~(\ref{eq:spin-spinor-contract}).

We often project spin components from our massive particles. As such, it would be useful to expand our spin-spinors in a basis of two-dimensional vectors $\zeta^{\pm I}$ in the $SU(2)$ little group space,
\begin{align}\label{eq:massive:spinor:expand:littlegroup}
      \b{\lambda}_\alpha^{\;\; I} &= \lambda_\alpha \zeta^{-I} + \eta_\alpha \zeta^{+I}\,\nonumber\\
  \tilde{\b{\lambda}}_{\dot\alpha I} &= \tilde\lambda_{\dot\alpha} \zeta^+_{I} + \tilde\eta_{\dot\alpha} \zeta^{-}_I\,,
\end{align}
Now, the massive $p_{\alpha\dot\alpha}$, from eq.~\eqref{eq:massive:momentum:spinor} takes the form,
\begin{align}\label{eq:massive:momentum:spinor2}
    p_{\alpha\dot\alpha}&= \lambda_{\alpha}\tilde\lambda_{\dot\alpha} \zeta^{-I}\zeta^+_{I} + \eta_{\alpha}\tilde\eta_{\dot\alpha}\zeta^{+I}\zeta^{-}_I \nonumber \\
    &= \lambda_{\alpha}\tilde\lambda_{\dot\alpha} - \eta_{\alpha}\tilde\eta_{\dot\alpha} 
\end{align}
where we have chosen $ \zeta^{-I}\zeta^+_{I} = - \zeta^{+I}\zeta^{-}_I = 1$, or explicity,
\begin{align}\label{eq:spin-spinor:expand}
  \zeta^+_I = \begin{pmatrix} 1 \\ 0 \end{pmatrix} \quad \text{and}\quad \zeta^-_I = \begin{pmatrix} 0 \\ 1 \end{pmatrix} \,.
\end{align}
Demanding the on-shell condition $\text{det}(p_{\alpha\dot\alpha}) = m^2$ imposes further constraints on our coefficients $\lambda$, $\tilde\lambda$, $\eta$ and $\tilde\eta$ in the little group space. To see this, we can again use $\text{det}(p_{\alpha\dot\alpha}) = \frac{1}{2}\epsilon^{\alpha\beta}\epsilon^{\dot\alpha\dot\beta}p_{\alpha\dot\alpha}p_{\beta\dot\beta}$, substitute eq.~\eqref{eq:massive:momentum:spinor2}, and get $m^2 = \la\lambda\eta\ra[\tilde\lambda\tilde\eta]$. Thus, our components in the little group space must satisfy
\begin{align}\label{eq:onshell:condition:massive}
    m =\la\lambda\eta\ra = [\tilde\lambda\tilde\eta]\,.
\end{align}
With these definitions, we have the following identities,
\begin{align}\label{eq:projections:identities}
  \lambda^{\alpha I}\zeta^+_I = \lambda^\alpha\,,
  \quad  \tilde{\lambda}^{\dot\alpha I} \zeta^+_I = \tilde\eta^{\dot\alpha}\,, \quad \lambda^{\alpha I}\zeta^-_I = - \eta^\alpha\,, \quad \tilde{\lambda}^{\dot\alpha I} \zeta^-_I = - \tilde\lambda^{\dot\alpha}\,,
\end{align}
which will be useful in taking the high energy limits.
Comparing the above to the helicity-spinors in eq.~\eqref{eq:helicity:def:spinor}, we see that the spin components in eq.~\eqref{eq:massive:spinor:expand:littlegroup} were constructed to be proportional their massless helicity counterparts. We can make the explicit connection to the massless spinors in eq.~\eqref{eq:massless:spinors:explicit} by 
\begin{align}
  \b{\lambda}_\alpha^{\;\; I} &= \lambda_\alpha \zeta^{-I} + \eta_\alpha \zeta^{+I}\,\nonumber\\
  &=\sqrt{E+p}\, \zeta^+_\alpha \zeta^{-I} + \sqrt{E-p}\, \zeta^-_\alpha \zeta^{+I}\,, \nonumber \\
  \tilde{\b{\lambda}}_{\dot\alpha I} &= \tilde\lambda_{\dot\alpha} \zeta^+_{I} + \tilde\eta_{\dot\alpha} \zeta^{-}_I\,\nonumber \\
  &= \sqrt{E+p}\, \tilde\zeta^-_{\dot\alpha} \zeta^+_{I} + \sqrt{E-p}\, \tilde\zeta^+_{\dot\alpha} \zeta^{-}_I\,,
\end{align}
where $\eta_\alpha$ and $\tilde\eta_{\dot\alpha}$ are also eigenstates of helicity and, \begin{align}
  \zeta^+_\alpha = \begin{pmatrix} c \\s \end{pmatrix}\,, \quad \zeta^-_\alpha = \begin{pmatrix} s^* \\ -c \end{pmatrix}\,, \quad \tilde\zeta^+_{\dot\alpha} = \begin{pmatrix} -s\\c \end{pmatrix}\,, \quad \tilde\zeta^-_{\dot\alpha} = \begin{pmatrix} c \\s^* \end{pmatrix}\,.
\end{align} 
Note that in the high energy limit, $E \gg m$, we have that $\sqrt{E+p}\rightarrow \sqrt{2E}$ and $\sqrt{E-p} \rightarrow m/\sqrt{2E}$. Thus, in this sense, the massive spin-spinors are constructed to coincide with their massless helicity spinors in the high energy limit.

\section{High Energy Limits\label{sec:HighEnergyLimits}}

In this section, we illustrate how to take the high energy limits of amplitudes with our conventions. We use the IR amplitude in eq.~\eqref{eq:Two:massive:spin:half:and:one:massive:spin:one}.
The kinematic part of the amplitude is given by,
\begin{align}\label{eq:appendix:massive:fermionamp}
  \c{A}(\b{1}^{1/2},\b{2}^{1/2},\b{3}^{1}) = X_1 \la\b{13}\ra[\b{32}] + X_2[\b{13}]\la\b{32}\ra + X_3 \la\b{13}\ra\la\b{32}\ra + X_4 [\b{13}][\b{32}]
\end{align}
which, after restoring the explicit $SU(2)$ little group indices takes the form,
\begin{align}
  \c{A}(\b{1}^{I},\b{2}^{J},\b{3}^{\{K_1,K_2\}}) &= X_1 \la\b{1}^I\b{3}^{\{K_1}\ra[\b{3}^{K_2\}}\b{2}^J] + X_2[\b{1}^I\b{3}^{\{K_1}]\la\b{3}^{K_2\}}\b{2}^J\ra \nonumber\\
  &+ X_3 \la\b{1}^I\b{3}^{\{K_1}\ra\la\b{3}^{K_2\}}\b{2}^J\ra + X_4 [\b{1}^I\b{3}^{\{K_1}][\b{3}^{K_2\}}\b{2}^J]
\end{align}
As previsouly mentioned, the momentum of a particle with spin-$s$ is described by a rank-$2s$ symmetric tensor, and has $2s+1$ dimensions. So the amplitude $\c{A}(\b{1}^{I},\b{2}^{J},\b{3}^{\{K_1,K_2\}})$ has dimension $2\times 2 \times 3 = 12$, corresponding to the different spin components of each particle. 
One can either take the high energy limit of the amplitude first, and attain all the $16$ spin projections, or project a particular spin configuration first, and then take the high energy limit. We will adopt the latter strategy.

Our goal is to determine the coefficient of the spin components $\c{A}(1^{+1/2},2^{+1/2},3^0)$ in the high energy limit of $\c{A}(\b{1}^{I},\b{2}^{J},\b{3}^{\{K_1,K_2\}})$. We begin by projecting the spin components, which is a straightfoward application of the identites in eq.~(\ref{eq:projections:identities}),
\begin{align}\label{eq:proj:example}
  \c{A}(\b{1}^{+1/2},\b{2}^{+1/2},\b{3}^0) &= \c{A}(\b{1}^{I},\b{2}^{J},\b{3}^{\{K_1,K_2\}}) \times \zeta^+_I \zeta^+_J (\zeta^+_{K_1}\zeta^-_{K_2} + \zeta^+_{K_2}\zeta^-_{K_1} ) \nonumber \\
  &= -X_1 \left( \la 13\ra[3\tilde\eta_2] + \la 1 \eta_3 \ra[\tilde\eta_3\tilde\eta_2]\right) - X_2 \left( [\tilde\eta_1 3]\la 32 \ra + [\tilde\eta_1\tilde\eta_3]\la\eta_32\ra \right) \nonumber \\
  &\hspace{12pt} -X_3 \left( \la13\ra\la\eta_32\ra + \la 1 \eta_3 \ra \la 32 \ra \right) - X_4 \left( [\tilde\eta_1 3][\tilde\eta_3\tilde\eta_2] + [\tilde\eta_1\tilde\eta_3][3\tilde\eta_2] \right)\,.
\end{align}

The second step is to take the high energy limit. This can be simplified by recalling that a three particle amplitude in four dimensions must have mass dimension $0$, and so the $X$'s must have mass dimension $-1$.  As such, terms $\c{O}(m^2)$ or higher will vanish in the high energy limit. In addition, $\eta,\tilde\eta \propto \sqrt{E-p} = m\sqrt{2E} + \c{O}(m^2)$ in the high energy limit. Thus, the only relevant terms are those that have at most one $\eta,\tilde\eta$. Thus, our next task is to convert terms involving $\eta$ or $\tilde\eta$ into a form that is suitable for taking a high energy limit. 

Any square brackets with $\tilde \eta$ can be converted to angle brackets through a Schouten identity $|1]\la1| - |\tilde\eta_1]\la\eta_1| + |2]\la2| - |\tilde\eta_2]\la\eta_2|+ |3]\la3|- |\tilde\eta_3]\la\eta_3| = 0$. For example, contracting with $[\tilde\eta_1|$, omitting terms of $\c{O}(m^2)$, and using the on-shell conditions in eq.~\eqref{eq:onshell:condition:massive}, we get $-m_1 \la1| + [\tilde\eta_12]\la2| + [\tilde\eta_23]\la3| = 0$, from which, either contracting with $|2\ra$ or $|3\ra$, we get $[\tilde\eta_12] =m_1 \frac{\la13\ra}{\la23\ra}$ and $[\tilde\eta_13] = m_1 \frac{\la12\ra}{\la32\ra}$. Repeated use of the Schouten identity, or applying a cyclic permutation on the set of identities $\left\{[\tilde\eta_11] = -m_1,[\tilde\eta_12] =m_1 \frac{\la13\ra}{\la23\ra}, [\tilde\eta_13] = m_1 \frac{\la12\ra}{\la32\ra}\right\}$ gives, for $\sum h>0$,
\begin{alignat}{6}\label{eq:he:replace:1}
  [\tilde\eta_11] &= -m_1\,, &\quad [\tilde\eta_12] &= m_1\frac{\la13\ra}{\la23\ra} \,, &\quad [\tilde\eta_13] &= m_1 \frac{\la12\ra}{\la32\ra} \,,\nonumber\\
  [\tilde\eta_22] &= -m_2\,, &\quad [\tilde\eta_23] &= m_2\frac{\la21\ra}{\la31\ra} \,, &\quad [\tilde\eta_21] &= m_2 \frac{\la23\ra}{\la13\ra}\, ,\\
  [\tilde\eta_33] &= -m_3\,, &\quad [\tilde\eta_31] &= m_3\frac{\la32\ra}{\la12\ra} \,, &\quad [\tilde\eta_32] &= m_3 \frac{\la31\ra}{\la21\ra}\,. \nonumber
\end{alignat}

Next, any angle brackets with $\eta$ can be written in a form more useful for a high energy limit by use of the on-shell condition eq.~\eqref{eq:onshell:condition:massive}. For example, if all that matters is $\la \eta_11\ra = -m_1$, then we can write $\la \eta_1| = -m_1 \frac{\la2|}{\la21\ra}$, from which we obtain $\la\eta_12\ra =0$ and $\la\eta_13\ra = -m_1 \frac{\la32\ra}{\la21\ra}$. Again, either by repeatedly using this, or applying a cyclic permutation on the set of identities $\left\{ \la\eta_11\ra = -m_1, \la\eta_12\ra =0, \la\eta_13\ra = -m_1\frac{\la32\ra}{\la21\ra}\right\}$, gives, for $\sum h >0$,
\begin{alignat}{6}\label{eq:he:replace:2}
  \la\eta_11\ra &= -m_1, &\quad \la\eta_12\ra &=0, &\quad \la\eta_13\ra &= -m_1\frac{\la32\ra}{\la21\ra} \,, \nonumber \\
  \la\eta_22\ra &= -m_2, &\quad \la\eta_23\ra &=0, &\quad \la\eta_21\ra &= -m_2\frac{\la13\ra}{\la32\ra}  \,,\\
  \la\eta_33\ra &= -m_3, &\quad \la\eta_31\ra &=0, &\quad \la\eta_32\ra &= -m_3\frac{\la21\ra}{\la13\ra} \nonumber\,. 
\end{alignat}
For $\sum h <0$, simply take eq.~(\ref{eq:he:replace:1}) and eq.~(\ref{eq:he:replace:2}) and replace all angle with square brackets and vice-versa.

Finally, we can take the high energy limit of eq.~(\ref{eq:proj:example}) by substituting eq.~(\ref{eq:he:replace:1}) and eq.~(\ref{eq:he:replace:2}), and dropping all terms $\c{O}(m^2)$ and higher,
\begin{align}
  \c{A}(\b{1}^{+1/2},\b{2}^{+1/2},\b{3}^0) &\overset{HE}{\longrightarrow} \c{A}({1}^{+1/2},{2}^{+1/2},{3}^0) = \la12\ra \left(X_1 m_2 - X_2 m_1 + X_3 m_3\right)
\end{align}
We summarize the results of this process on all spin components of the amplitude in eq.~\eqref{eq:appendix:massive:fermionamp} in Table~\ref{fermionmass:tab:HE1}.
  \begin{table}[!h]
  \centering
   \begin{tabular}{| c |c | c | c |c | c | c | c |}
      \hline
      1 & 2 & 3 & $\sum h$ & $\ \la\b{13}\ra[\b{32}]$ & $ [\b{13}]\la\b{32}\ra$ & $ \la\b{13}\ra\la \b{32}\ra $ & $ [\b{13}][\b{32}]$ \\\hline
       &  &  & & $X_1$ & $X_2$ & $X_3$ & $X_4$ \\
      \hline \hline
      $+1/2$ & $+1/2$ & $+1$ & $+2$ & $ 0$ & $ 0 $ & $ \la 13\ra\la32\ra$ & $ 0$\\ \hline
      $-1/2$ & $+1/2$ & $+1$ & $+1$ & $ 0$ & $m_3\frac{\la32\ra^2}{\la12\ra} $ & $ m_3\frac{\la32\ra^2}{\la12\ra} $ & $0 $ \\ \hline
      $+1/2$ & $-1/2$ & $+1$ & $+1$ & $-m_3 \frac{\la13\ra^2}{\la12\ra} $ & $ 0 $ & $ 0$ & $0 $\\ \hline
      $-1/2$ & $-1/2$ & $+1$ & $0$ & $ 0$ & $0$ & $0 $ & $ 0$\\ 
      \hline\hline
      $+1/2$ & $+1/2$ & $-1$ & $0$ & $0 $ & $ 0$ & $ 0$ & $ 0$\\ \hline
      $-1/2$ & $+1/2$ & $-1$ & $-1$ &  $ 0$ & $ -m_3 \frac{[13]^2}{[12]} $ & $0$ & $ 0 $\\ \hline
      $+1/2$ & $-1/2$ & $-1$ & $-1$ & $m_3 \frac{[23]^2}{[12]} $ & $0 $ & $ 0$ & $ m_3 \frac{[23]^2}{[12]}$\\ \hline
      $-1/2$ & $-1/2$ & $-1$ & $-2$ & $ 0$ & $ 0$ & $ 0$ & $[13][32]$\\ 
      \hline \hline
      $+1/2$ & $+1/2$ & $0$  & $+1$ & $ m_2\la12\ra $ & $ -m_1\la12\ra $ & $ m_3\la12\ra $ & $ 0$\\ \hline
      $-1/2$ & $+1/2$ & $0$  & $0$ & $ 0 $ & $0 $ & $ 0$ & $ 0$\\ \hline
      $+1/2$ & $-1/2$ & $0$  & $0$ & $ 0$ & $ 0 $ & $ 0 $ & $0$\\ \hline
      $-1/2$ & $-1/2$ & $0$  & $-1$ & $ -m_1 [12] $ & $ m_2 [12] $ & $ 0 $ & $m_3[12] $\\ \hline
   \end{tabular}\caption{High energy limits of two massive spin-1/2, one massive spin-1\label{fermionmass:tab:HE1}}
\end{table}

\section{Representation of the Higgs \label{appendix:repofHiggs}}

We would like to construct a representation that makes the four degrees of freedom clear. This can be done, for example, by considering a four dimensional real representation of $SO(4)$ and then extracting an $SU(2) \times U(1)$ subgroup \cite{Bachu:2019ehv,Helset:2018fgq}. We will instead take the approach \cite{Langacker:2017uah}.

The scalar field has four real degrees of freedom, $\phi_1$, $\phi_2$, $\phi_3$ and $\phi_4$, and are usually packaged in the form
\begin{align}
    \phi = \frac{1}{\sqrt{2}} \begin{pmatrix}
        \phi_1 + i \phi_2 \\
        \phi_3 + i \phi_4
    \end{pmatrix}
\end{align}
To make contact with a Lagrangian formalism, the $SU(2)_L\times U(1)_Y$ covariant derivative acting on $\phi$ is given by,
\begin{align}
D_\mu \phi_i = \partial_\mu \phi_i + W^k_\mu (\mathcal{L}^k)_i^{\:j} \phi_j + B_\mu (\mathcal{L}^4)_i^{\:j} \phi_j
\end{align}
where $i,j = 1,2$,  $k=1,2,3$ and $(\mathcal{L}^k)_i^{\:j} = \frac{1}{2} i g_1 (\tau^k)_i^{\:j}$ and $(\mathcal{L}^4)_i^{\:j} = \frac{1}{2} i g_2 Y \delta_i^{\:j}$
. To see how the real degrees of freedom transform, we can first consider a reducible scalar representation $\Phi$, a complex 4-component vector,
\begin{align}
    \Phi = \begin{pmatrix} \phi \\ \phi^\dagger \end{pmatrix}  = \frac{1}{\sqrt{2}} \begin{pmatrix} \phi_1 + i \phi_2 \\ \phi_3 + i \phi_4 \\ \phi_1 - i \phi_2 \\ \phi_3 -i \phi_4 \end{pmatrix}= \begin{pmatrix} \phi^+ \\ \phi^0 \\ \phi^- \\ \phi^{0*} \end{pmatrix}
\end{align}
with half the components redundant. 
Now, $\Phi$ transforms under the reducible representation,
\begin{align}
\mathcal{L}^k_\Phi &= \begin{pmatrix} \mathcal{L}^k_\phi & \\
& - \mathcal{L}_\phi^{k*} \end{pmatrix} \,.
\end{align}
Since $\mathcal{L}^3_\Phi$ and $\mathcal{L}^4_\Phi$ are diagonal, that is,
\begin{align}
\mathcal{L}^3_\Phi = \frac{1}{2}ig_1 \begin{pmatrix}\sigma^3 & \\ & -\sigma^3\end{pmatrix} \, \quad \text{and} \quad \mathcal{L}^4_\Phi = \frac{1}{2}ig_2 Y \begin{pmatrix}\delta & \\ & -\delta\end{pmatrix}
\end{align}
we can read of the $z$-component of the $SU(2)_L$ spin, as well as the $U(1)_Y$ charge of every component of the scalar field.
  \begin{table}[!h]
  \centering
   \begin{tabular}{| c |c | c | c |}
      \hline
      Component & $\c{L}^3$ & $\c{L}^4$ & $Q$ \\ 
      \hline 
      $\phi^+$ & $+\frac{1}{2}g_1$&$+\frac{1}{2}g_2 Y$ & $+e$ \\ \hline
      $\phi^0$ & $-\frac{1}{2}g_1$&$+\frac{1}{2}g_2Y$ & $0$ \\ \hline
      $\phi^-$ & $+\frac{1}{2}g_1$&$-\frac{1}{2}g_2Y$ & $-e$ \\ \hline
      $\phi^{0*}$ & $-\frac{1}{2}g_1$&$-\frac{1}{2}g_2Y$ & $0$ \\ \hline
   \end{tabular}\caption{Higgs field}
\end{table}

The usefulness of this reducible representation is that we can now perform a simple change of basis to act on the real degrees of freedom $\phi_h$. The two representations are related by $\Phi = A \phi_h$, with
\begin{align}
    A = \frac{1}{\sqrt{2}} \begin{pmatrix} \mathbf{1} & i\mathbf{1} \\ \mathbf{1} & -i \mathbf{1} \end{pmatrix} \quad \text{and} \quad \phi_h = \begin{pmatrix}\phi_1 \\ \phi_3 \\ \phi_2 \\ \phi_4 \end{pmatrix} \,.
\end{align}
Thus, in this basis, the representation matrices are given by $\mathcal{L}^i_{\phi_h}=A^\dagger\mathcal{L}^i_\Phi A$.

Explicitly, we have
\begin{alignat}{4}
\c{L}^1_\Phi &= \frac{g_1}{2}\left(
\begin{array}{cccc}
 0 & 1 & 0 & 0 \\
 1 & 0 & 0 & 0 \\
 0 & 0 & 0 & -1 \\
 0 & 0 & -1 & 0 \\
\end{array}
\right)\,,&\quad \c{L}^2_\Phi &= \frac{g_1}{2} \left(
\begin{array}{cccc}
 0 & -i & 0 & 0 \\
 i & 0 & 0 & 0 \\
 0 & 0 & 0 & -i \\
 0 & 0 & i & 0 \\
\end{array}
\right), \nonumber
\\
\c{L}^3_\Phi &= \frac{g_1}{2}\left(
\begin{array}{cccc}
 1 & 0 & 0 & 0 \\
 0 & -1 & 0 & 0 \\
 0 & 0 & -1 & 0 \\
 0 & 0 & 0 & 1 \\
\end{array}
\right),&\quad
\c{L}^4_\Phi &= \frac{g_2}{2} Y \left(
\begin{array}{cccc}
 1 & 0 & 0 & 0 \\
 0 & 1 & 0 & 0 \\
 0 & 0 & -1 & 0 \\
 0 & 0 & 0 & -1 \\
\end{array}
\right)\,,
\end{alignat}
in the basis $\phi^+,\phi^0,\phi^-,\phi^{0*}$ and 
\begin{alignat}{4} \label{eq:higgs:so4:origin}
   \c{L}^1_{\phi_h} &= \frac{g_1}{2}\left(
\begin{array}{cccc}
 0 & 0 & 0 & i \\
 0 & 0 & i & 0 \\
 0 & -i & 0 & 0 \\
 -i & 0 & 0 & 0 \\
\end{array}
\right),&\quad
\c{L}^2_{\phi_h} &=\frac{g_1}{2}\left(
\begin{array}{cccc}
 0 & -i & 0 & 0 \\
 i & 0 & 0 & 0 \\
 0 & 0 & 0 & -i \\
 0 & 0 & i & 0 \\
\end{array}
\right) \,,\nonumber\\
\c{L}^3_{\phi_h} &=\frac{g_1}{2}\left(
\begin{array}{cccc}
 0 & 0 & i & 0 \\
 0 & 0 & 0 & -i \\
 -i & 0 & 0 & 0 \\
 0 & i & 0 & 0 \\
\end{array}
\right),&\quad
\c{L}^4_{\phi_h} &=\frac{g_1}{2}Y\left(
\begin{array}{cccc}
 0 & 0 & i & 0 \\
 0 & 0 & 0 & i \\
 -i & 0 & 0 & 0 \\
 0 & -i & 0 & 0 \\
\end{array}
\right)\,.
\end{alignat}
in the basis $\phi_1,\phi_2,\phi_3,\phi_4$.
Eq.~\eqref{eq:higgs:so4:origin} corresponds to the subset of $SO(4)$ generators usually used to describe the real degrees of freedom of the Higgs.

\bibliography{sample}
\bibliographystyle{JHEP}

\end{document}